%% file: RaulRPrado_NA61.tex
\documentclass[epj,twocolumn]{webofc}
\usepackage[varg]{txfonts}   % Web of Conferences font
\woctitle{ISVHECRI 2018}

\usepackage{framed,color}
\usepackage{xspace}
\usepackage{stmaryrd}
\usepackage{rviewport}
\usepackage{comment}
\usepackage{overpic}
\usepackage{amsbsy}
\usepackage{multirow}
\usepackage{cleveref}
\crefname{chapter}{Chap.}{Chaps.}
\crefname{appendix}{App.}{Apps.}
\crefname{section}{Sec.}{Secs.}
\crefname{paragraph}{Sec.}{Secs.}
\crefname{table}{Tab.}{Tabs.}
\crefname{figure}{Fig.}{Figs.}
\crefname{equation}{Eq.}{Eqs.}
\crefname{item}{item}{items}
\usepackage[font=small]{caption}

\usepackage{lineno}
\input{TexAbbreviations}

\input{ThisAbbreviations}

\include{bib_aliases}

\begin{document}

\title{Recent results from the cosmic ray program of the \NASixtyOne experiment}

\author{\firstname{Raul R.} \lastname{Prado}\inst{1,2,3}\fnsep\thanks{\email{raul.prado@desy.de}}
  for the \NASixtyOne Collaboration\thanks{Full author list: http://shine.web.cern.ch/content/author-list} }
\institute{Deutsches Elektronen-Synchrotron (DESY), Platanenallee 6, D-15738 Zeuthen, Germany
  \and
  IKP, Karlsruhe Institute of Technology (KIT), Postfach 3640, D-76021 Karlsruhe, Germany
  \and
  Instituto de F\'isica de S\~ao Carlos (IFSC/USP), Av. Trabalhador S\~ao-carlense 400, 13566-590, S\~ao Carlos, Brazil  
}

\abstract{%
  \NASixtyOne is a fixed target experiment designed to study
  hadron-proton, hadron-nucleus and nucleus-nucleus interactions at the
  CERN Super-Proton-Synchrotron.  In this paper we summarize
  the results from pion-carbon collisions recorded at beam momenta of 158
  and 350 \GeVc.  Hadron production measurements in this type of
  interactions is of fundamental importance for the understanding of the
  muon production in extensive air showers. In particular, production of
  (anti)baryons and \r0 are mechanisms responsible for increasing the
  number of muons which reaches the ground. The underestimation of the
  (anti)baryons or \r0 production rates in current hadronic interaction
  models could be one of the sources of the excess of muons observed by
  cosmic ray experiments. The results on
  the production spectra of \pions, \kaons, \proton, \antiproton,
  \lamb, \antilamb, \kzeros, \r0, $\omega$ and \kstar are presented, as well as
  their comparison to predictions of hadronic interaction models
  currently used in air shower simulations.
}
\maketitle
%
%%%%%%%%%%%%%%%%%%%%%%%%%%%%%%%%
\section{Introduction}
\label{sec:intro}

Measurements of ultrahigh energy cosmic rays are only possible
through the detection of secondary particles produced in extensive air showers (EAS).
The inference of some of the properties of the primary cosmic ray particles
like their nuclear mass, relies on the comparison of measured EAS observables
to predictions from simulations~\cite{Kampert:2012mx}.
These simulations are performed by Monte Carlo codes
that make use of hadronic interaction models to describe the nucleus-air and hadron-air
collisions along the shower development~\cite{Engel:2011zzb}.
Although simulations using recent hadronic models
can provide a good overall description of EAS,
it has been observed by cosmic ray experiments that
the hadronic models fail on describing the muon production in EAS.
Measurements by HiRes-MIA~\cite{\HiResMIAMuonPaper},
Pierre Auger Observatory~\cite{\AugerHASMuonPaper,\AugerTopDownPaper,\AugerMPDPaper,\AugerDeltaPaper},
Telescope Array~\cite{Abbasi:2018fkz},
KASCADE-Grande~\cite{Apel:2017thr},
IceTop/IceCube~\cite{\IceTopMuonPaper}
and Sugar~\cite{\SugarMuonPaper} show that
there is an inconsistency between data and simulations for 
observables related to the muonic component of air showers. 
In particular,
the number of muons (\nmu) obtained from simulations is observed to be
significantly smaller than the measured ones, which is known 
as the ``muon deficit problem''.

The majority of muons in EAS are produced by the decay of
charged mesons, which in turn, are produced in meson-air and nucleon-air
interactions. Depending on the primary energy
and detection distance, the relevant meson-air and nucleon-air interaction energies
are between 10 and 1000 \GeV~\cite{Meurer:2005dt,\IoanaICRC}.
Therefore, measurements of particle production
in this energy range are of great value for understanding muon production
in EAS and consequently for improving its modeling. Of particular interest
are the production spectra of (anti-)baryons and \r0 in meson-air and nucleon-air.
It is well known ~\cite{\EposPaper,Drescher:2007hc} 
that the production of (anti-)baryons and \r0
mesons in hadronic interactions is important to predict the muon
content of air showers. Therefore the production cross sections
of these particles needs to be known accurately for a precise modeling
of air showers

\NASixtyOne experiment~\cite{\NASixtyOnePaper} (see~\cref{sec:shine})
has provided a number of particle production and cross section
measurements which are relevant for
the modeling of hadronic interaction in EAS (e.g. Refs~\cite{Abgrall:2015hmv,Aduszkiewicz:2017sei}).
In this paper, however, the focus will be on the results of the 2009 run with negatively
charged pion beam colliding against a thin carbon target (\piC data) at 158 and 350 \GeVc.
Since $\pi$-air is the most abundant hadronic interaction occurring in an EAS,
our \piC data is of high relevance for the tuning of 
hadronic interaction models
dedicated to EAS simulations.
The results are presented in three parts: the spectra of charged hadrons
(\pions, \kaons, \proton and \antiproton)
are presented in~\cref{sec:hadrons}, the spectra of \vzero mesons
(\lamb, \antilamb and \kzeros) in~\cref{sec:vzero}
and the spectra of resonance mesons (\r0, \kstar and $\omega$) in \cref{sec:resonance}.

%%%%%%%%%%%%%%%%%%%%%%%%%%%%%%%%
\section{The \NASixtyOne experiment}
\label{sec:shine}

\begin{figure}
  \centering
  \begin{overpic}[clip, rviewport=0 0 1 1,width=0.495\textwidth]{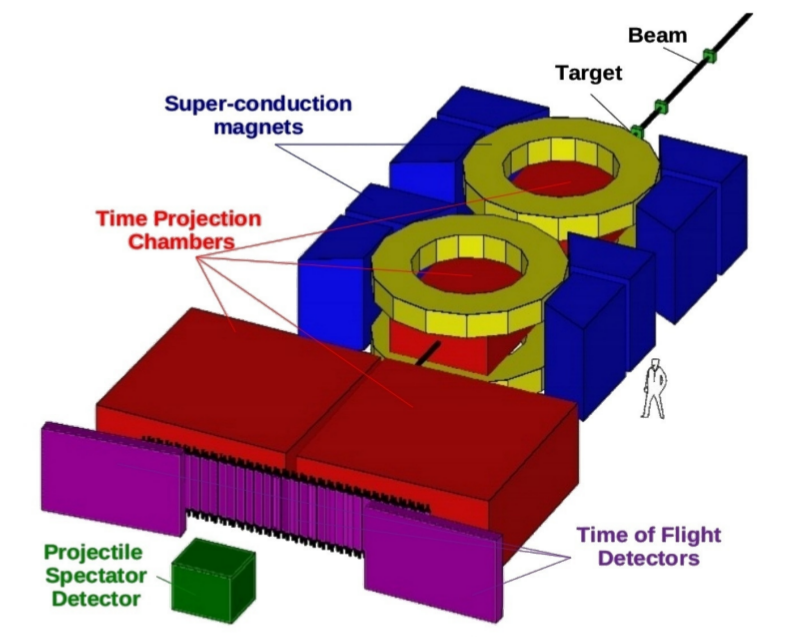}
    \put(15.5,23){}
  \end{overpic}
  \caption{Schematic layout of the \NASixtyOne experiment~\cite{\NASixtyOnePaper}.}
  \label{fig:na61}
\end{figure}

\NASixtyOne (SHINE = SPS Heavy Ion and Neutrino Experiment)
is a fixed target experiment at the CERN SPS designed to study
hadron production in nucleus-nucleus and hadron-nucleus
collisions. Its physics goals comprise a) the strong interaction
program, which investigates the properties of the onset of
deconfinement and search for the critical point of strongly
interacting matter, b) the neutrino program,
to precisely measure the hadron production important to calculate
the neutrinos and antineutrino fluxes in the T2K neutrino experiment~\cite{\T2KPaper},
and c) the cosmic rays program, focused on the measurements of the
hadron and meson production which are most relevant for the modeling
of extensive air showers. The full description of the \NASixtyOne experiment
and its science program can be found in Ref.~\cite{\NASixtyOnePaper}

The \NASixtyOne detector measures charged particles produced
by the collision of the beam particles with the target through
a set of five Time Projection Chambers (TPC). Since two of the TPCs
are placed in the magnetic field produced by superconducting
dipole magnets, the charge and the momenta of the particles
can be measured and the achieved resolution on \pp is of the order of
$\sigma(p)/p^2 = 10^{-4}$ (\GeVc)$^{-1}$. Additionally, the
energy loss per unit of length (\dedx) in the TPCs is used in this
work for particle identification. The experimental layout of the
\NASixtyOne detector is shown in~\cref{fig:na61}.

A beam detector system composed of scintillation and Cherenkov counters is
placed upstream of the detector to identify and measure the
beam particles. The position of the beam is measured by a
set of three beam position detectors, which are also placed upstream of
the target.

%%%%%%%%%%%%%%%%%%%%%%%%%%%%%%%%
\section{Production of \pions, \kaons, \proton and \antiproton}
\label{sec:hadrons}

\begin{figure}
  \centering
  \begin{overpic}[clip, rviewport=0 0 1 1,width=0.4\textwidth]{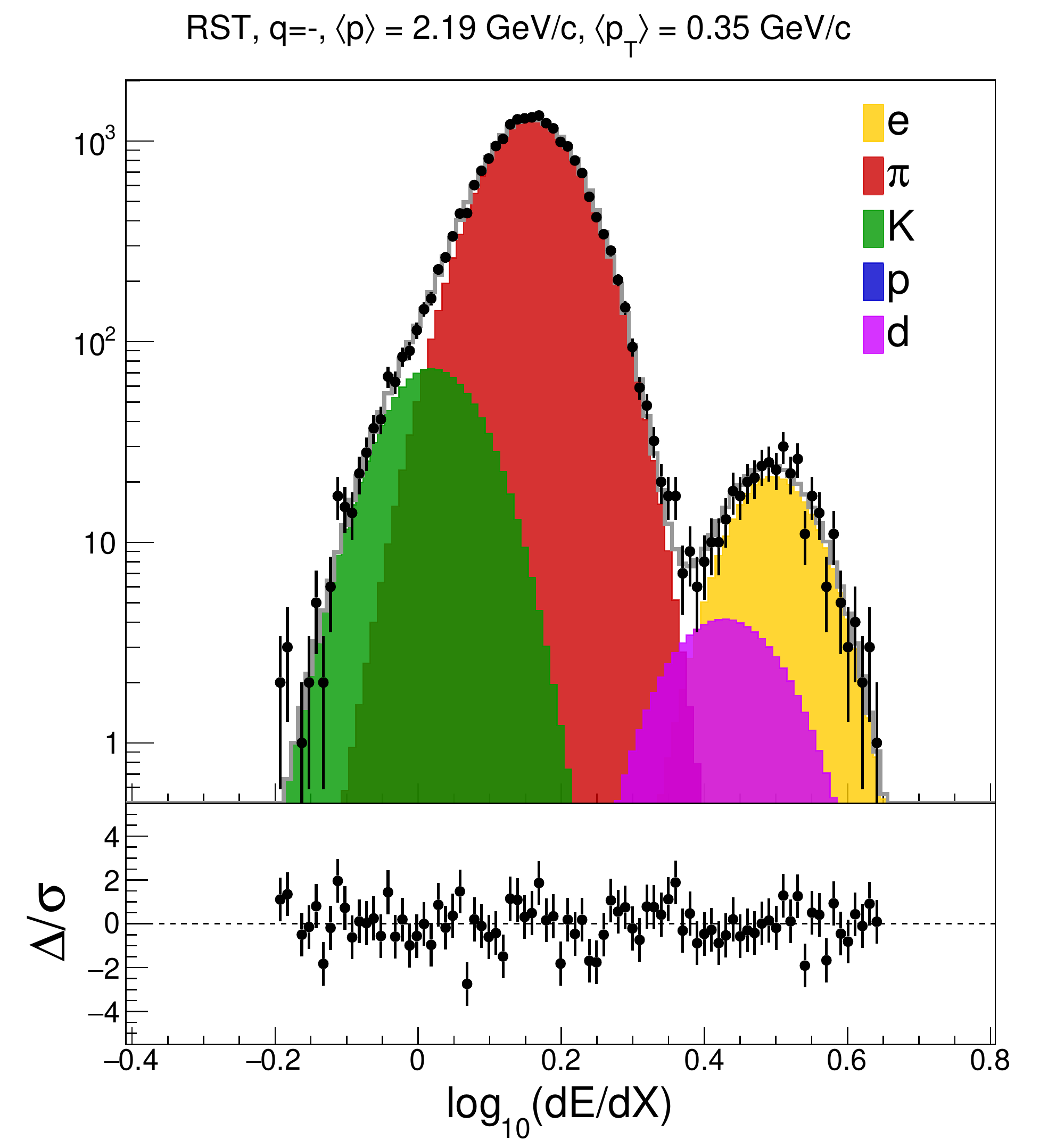}
  \end{overpic}

  \vspace{0.3cm}
  
  \begin{overpic}[clip, rviewport=0 0 1 1,width=0.4\textwidth]{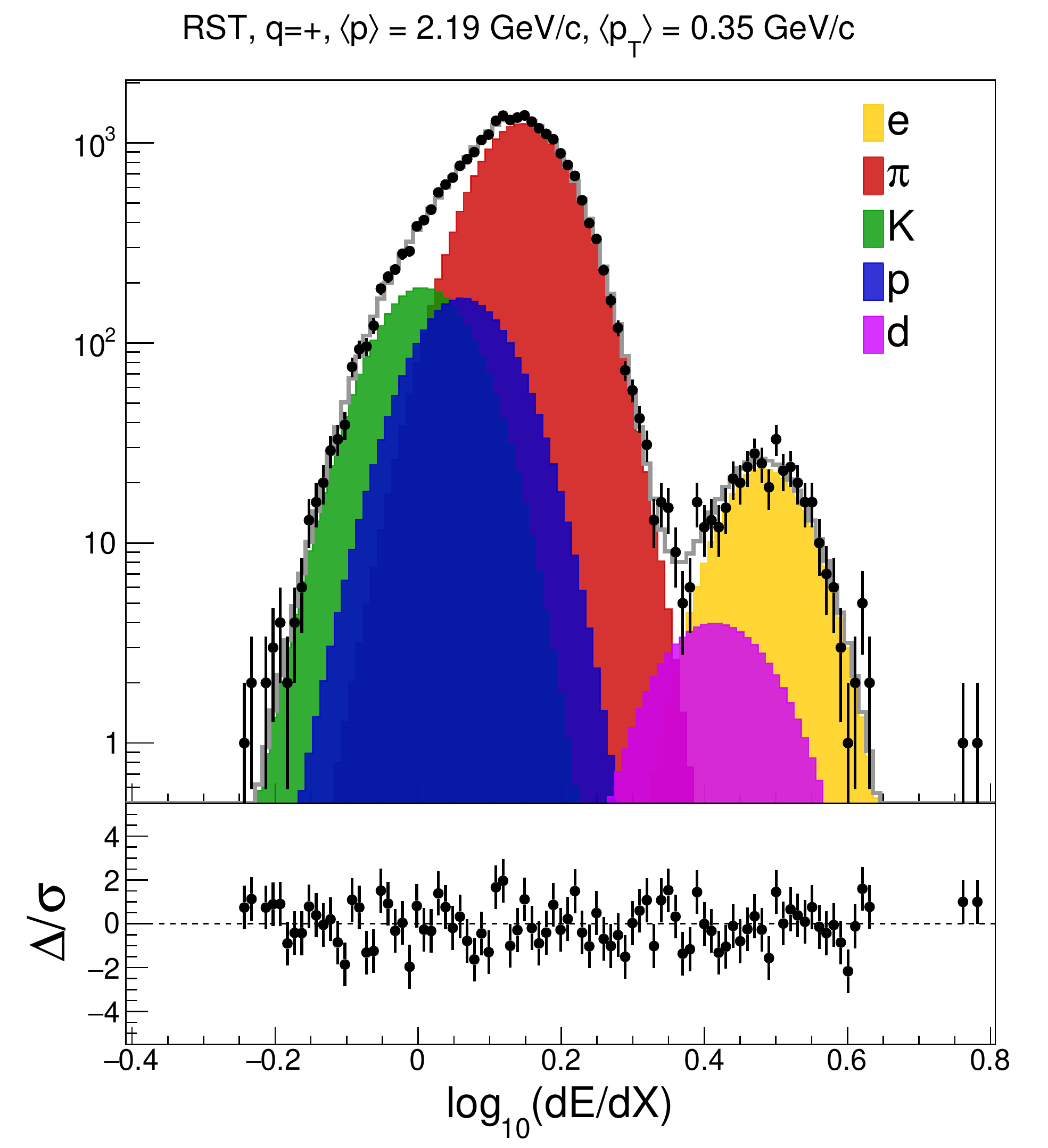}
    \put(37,58){}
  \end{overpic}

  \caption{Example of the \dedx distributions for one phase space
    bin ($\langle p \rangle = 2.19 \GeVc$ and $\langle p_\text{T}\rangle = 0.35 \GeVc$)
    of the 158 \GeVc data set.
    The black markers show the measured distributions and the colored
    distributions show the result of the \dedx fit. Negatively charged
    particles are shown on the top and positively ones on the bottom.}
  \label{fig:hadron:dedx}
\end{figure}

Charged particles are identified in \NASixtyOne
by the track-by-track measurement of the deposited energy,
\dedx, performed by the TPCs.
After splitting the data into bins of total and transverse momentum
(\pp and \pT), a \dedx model is fitted to the measured \dedx
distributions by accounting for contributions of 5 particle types
($e$, $\pi$, $K$, $p$ and deuterons). From the results of the fit,
the particle yields of \pions, \kaons and \protons are determined.
Examples of measured \dedx distributions and of the results of the
\dedx fit are shown in~\cref{fig:hadron:dedx}. After performing
the particle identification through the \dedx fit, the detector effects
(e.g. acceptance, efficiency) are corrected by using a set of Monte Carlo
simulations and the spectra are derived. A more detailed description
of the analysis procedure can be found in Ref.~\cite{\RaulICRC}.

The single-differential spectra
as a function of \pp (integrated over \pT) for
\pions, \kaons and \protons are shown
in~\cref{fig:hadron:int158,fig:hadron:int350},
where the measurements are compared to the predictions of
\EposLong~\cite{\EposPaper}, 
\SibyllLong~\cite{\SibyllPaper}, \SibyllNewLong~\cite{\NewSibyllCPaper},
\QGSJetLong~\cite{\QGSJetPaper} and \EposLHCLong~\cite{\EposLHCPaper}.
The double-differential spectra as a function of \pp and \pT can
be found in Ref.~\cite{\AlexICRC} for the \pions spectra and in Ref.~\cite{\RaulICRC} for the \kaons and \protons spectra. 

\begin{figure*}[!h]
  \centering

  \begin{overpic}[clip, rviewport=0 0 1 1,width=0.33\textwidth]{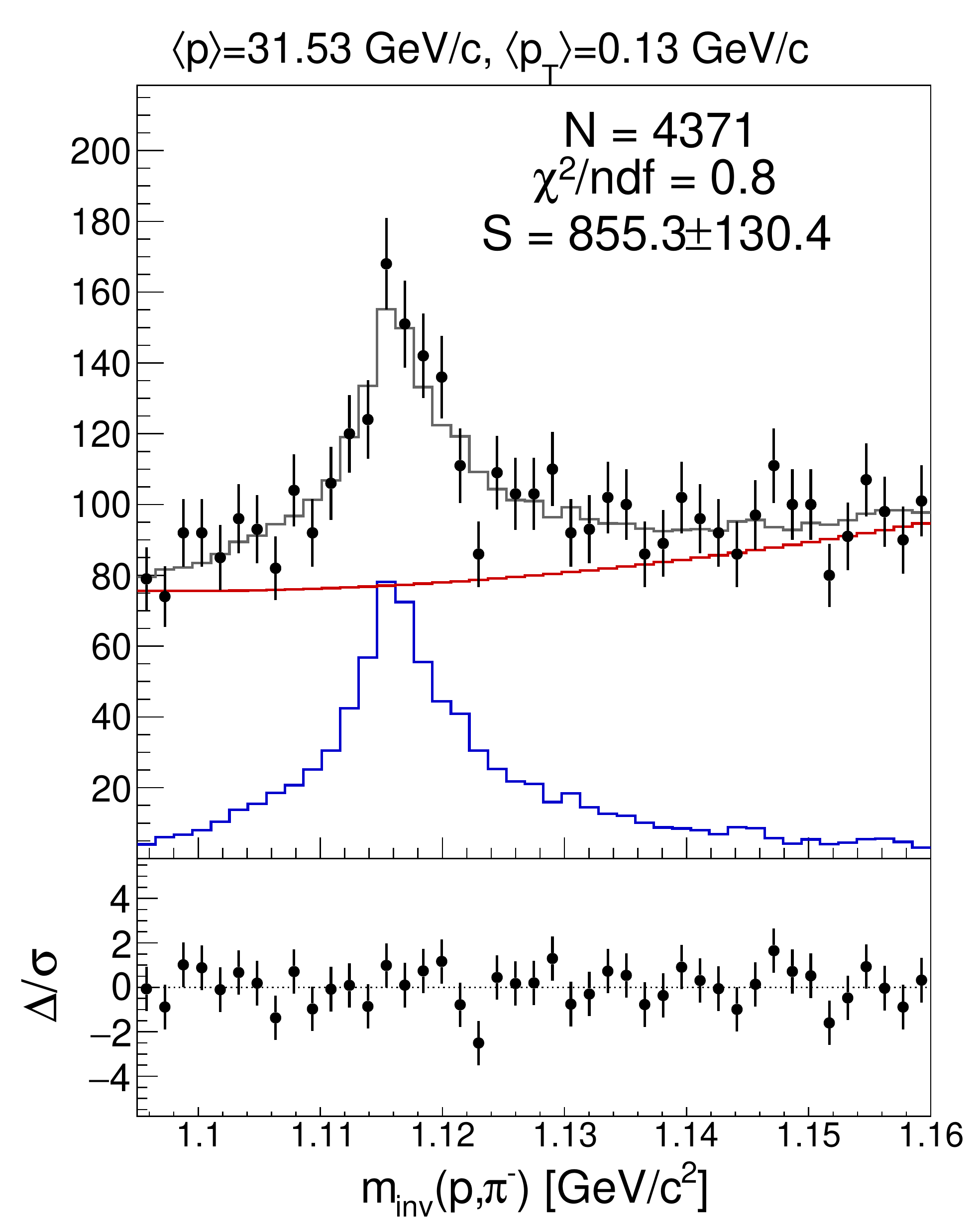}
    \put(17,82){\huge\lamb}
  \end{overpic}
  \begin{overpic}[clip, rviewport=0 0 1 1,width=0.33\textwidth]{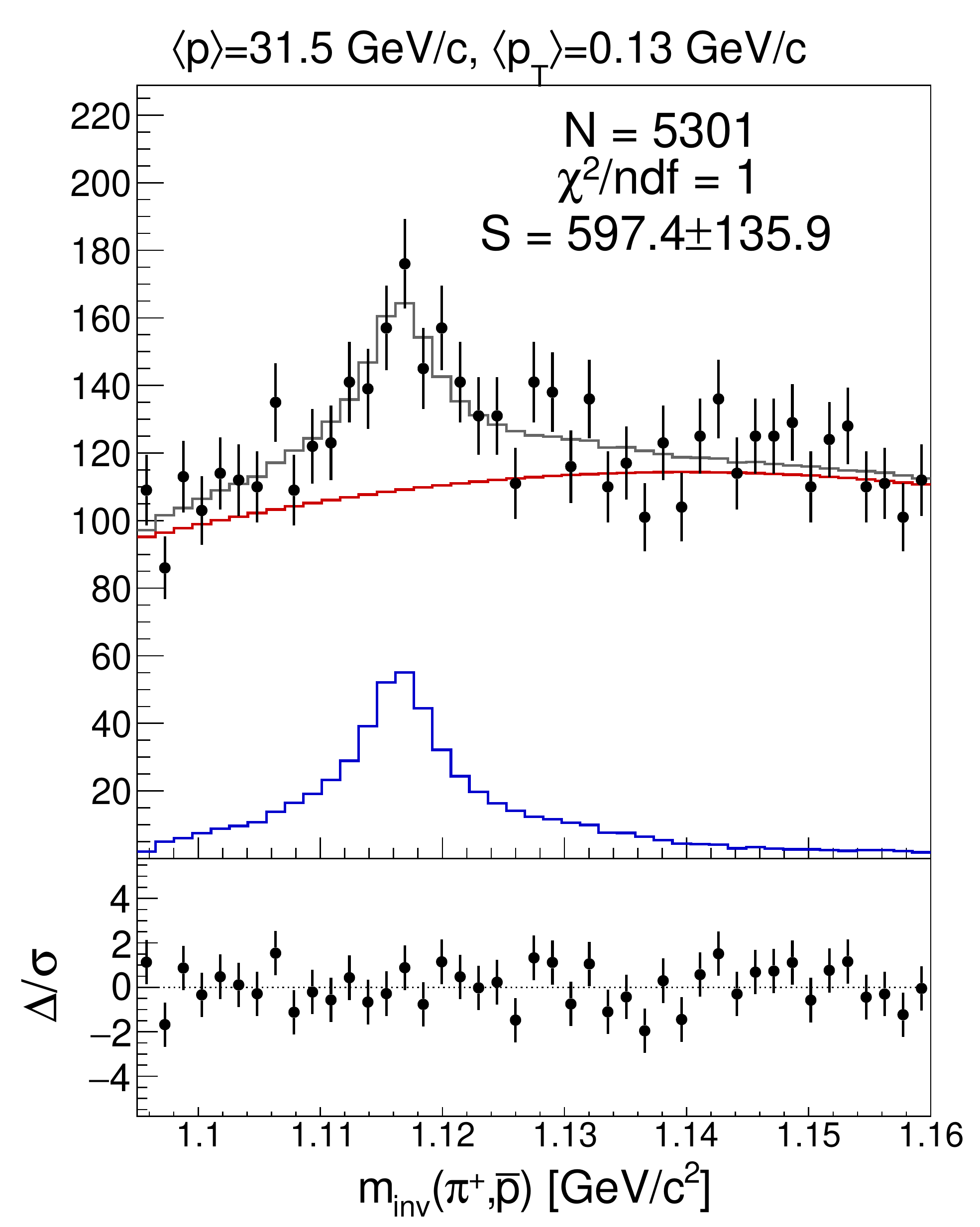}
    \put(17,82){\huge\antilamb}
  \end{overpic}
  \begin{overpic}[clip, rviewport=0 0 1 1,width=0.33\textwidth]{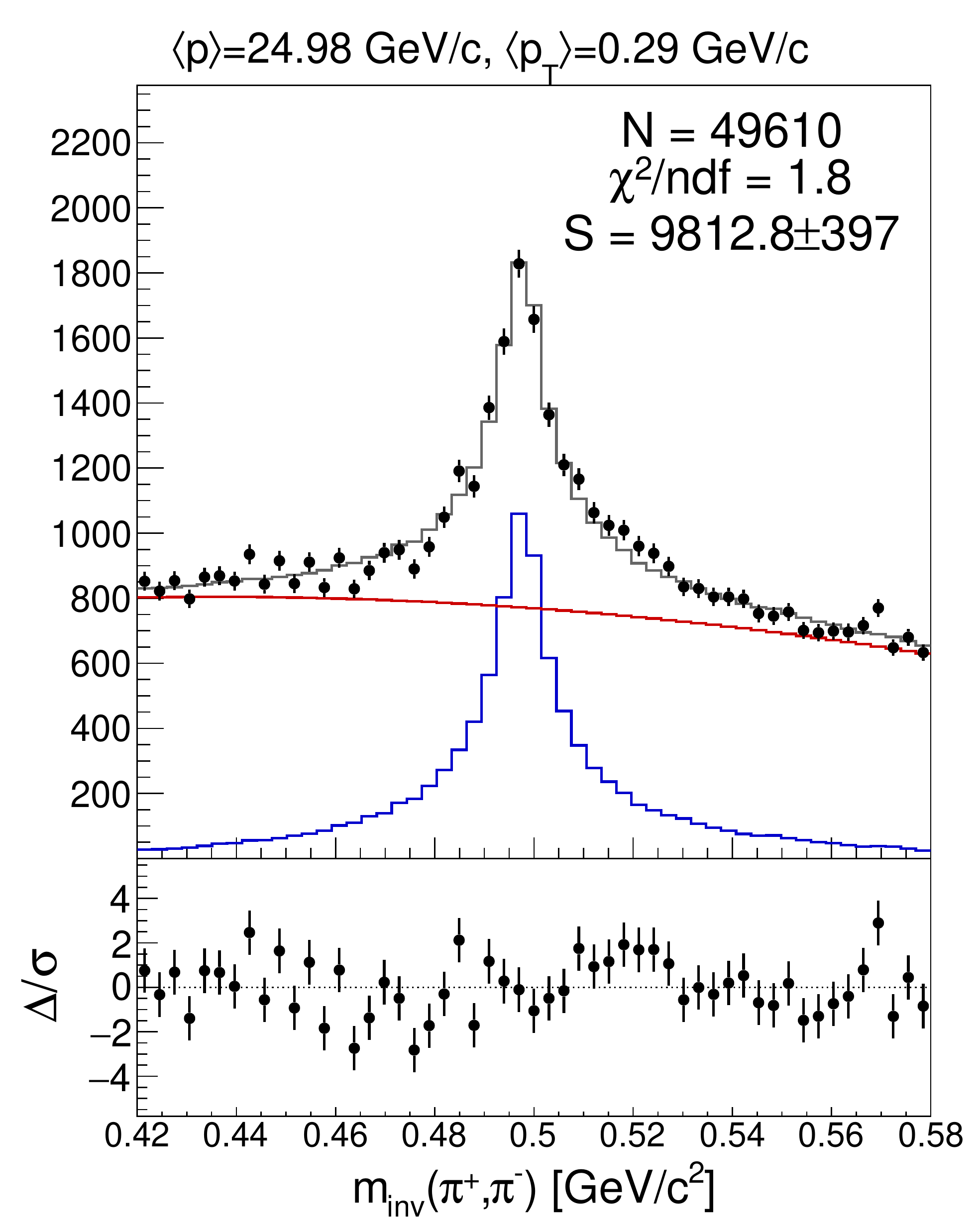}
    \put(17,82){\huge\kzeros}
  \end{overpic}

  \caption{Example of the \minv distributions for one phase space
    bin ($\langle p \rangle$ and $\langle p_\text{T}\rangle$ are indicated on
    the top of each plot) of the 158 \GeVc data set.
    The black markers show the measured distributions and the colored
    lines show the result of the signal extraction fit, where the signal
    is shown in blue and the background in red.}
  \label{fig:vzero:mass}
\end{figure*}

%%%%%%%%%%%%%%%%%%%%%%%%%%%%%%%%
\section{Production of \lamb, \antilamb and \kzeros}
\label{sec:vzero}

Since \lambs and \kzeros are neutral weakly decaying particles,
they can be measured by \NASixtyOne through the detection of the
charged particles which are produced in their decays.
The invariant mass (\minv) spectra for a given decay channel
can then be used to extract their signal.
The decay channels used here are $\Lambda\rightarrow p + \pi^-$,
$\bar{\Lambda}\rightarrow \bar{p} + \pi^+$,
$K_\text{S}^0\rightarrow \pi^+ + \pi^-$.
To extract the signal, the \minv distributions were fitted
by considering a signal contribution, modeled by using Monte Carlo templates,
and the background, modeled by a 2nd-degree polynomial function.
Examples of the fitted \minv distributions are shown in~\cref{fig:vzero:mass}.

\begin{figure*}
  \centering

  \begin{overpic}[clip, rviewport=0 0 1 1,width=0.495\textwidth]{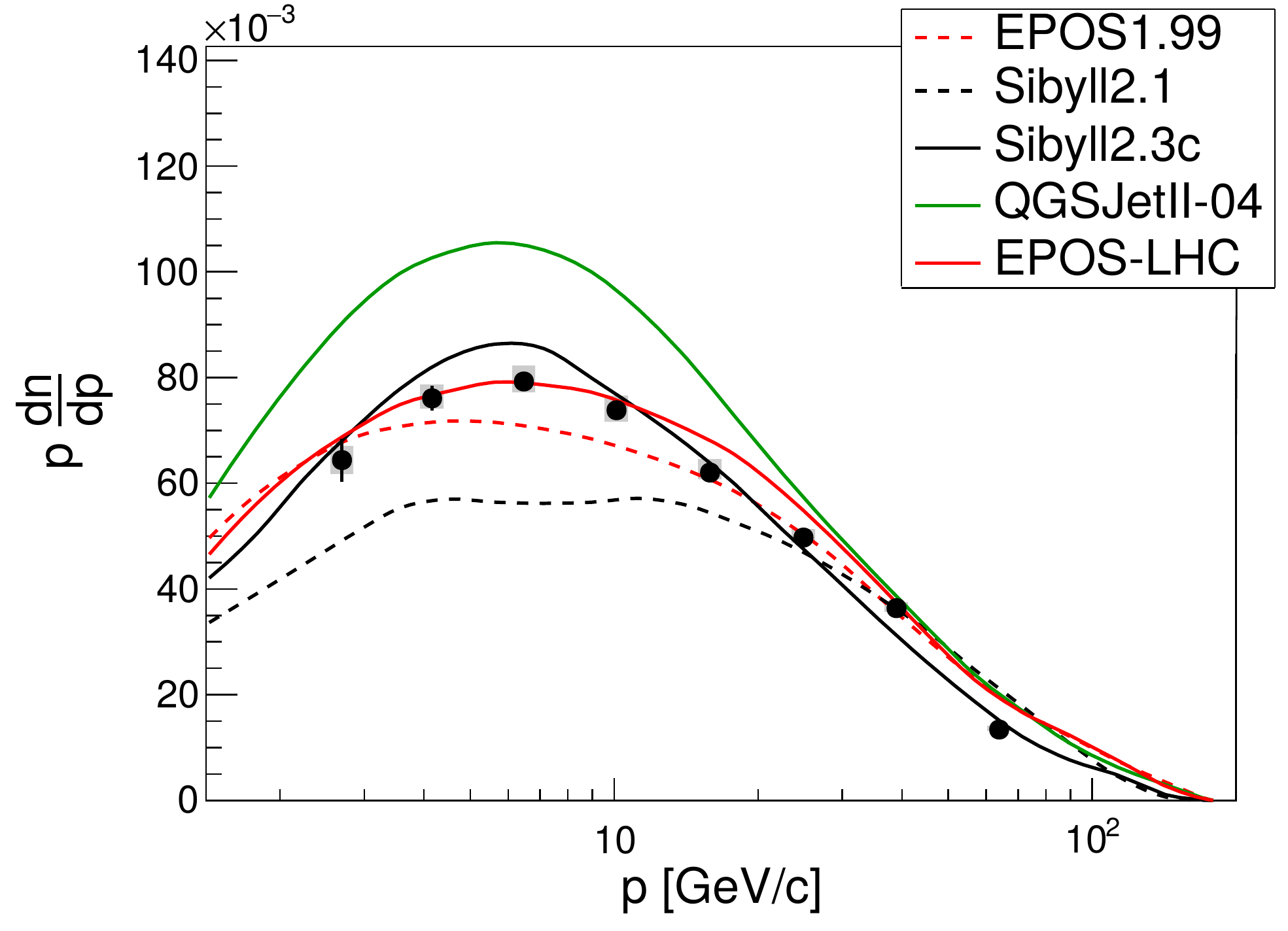}
    \put(70,75){\Large\NASixtyOne PRELIMINARY}
    \put(28,63){$\pi^-$+C$\rightarrow$\kzeros+X}
    \put(29,58){at 158 \GeVc}
  \end{overpic}
  \begin{overpic}[clip, rviewport=0 0 1 1,width=0.495\textwidth]{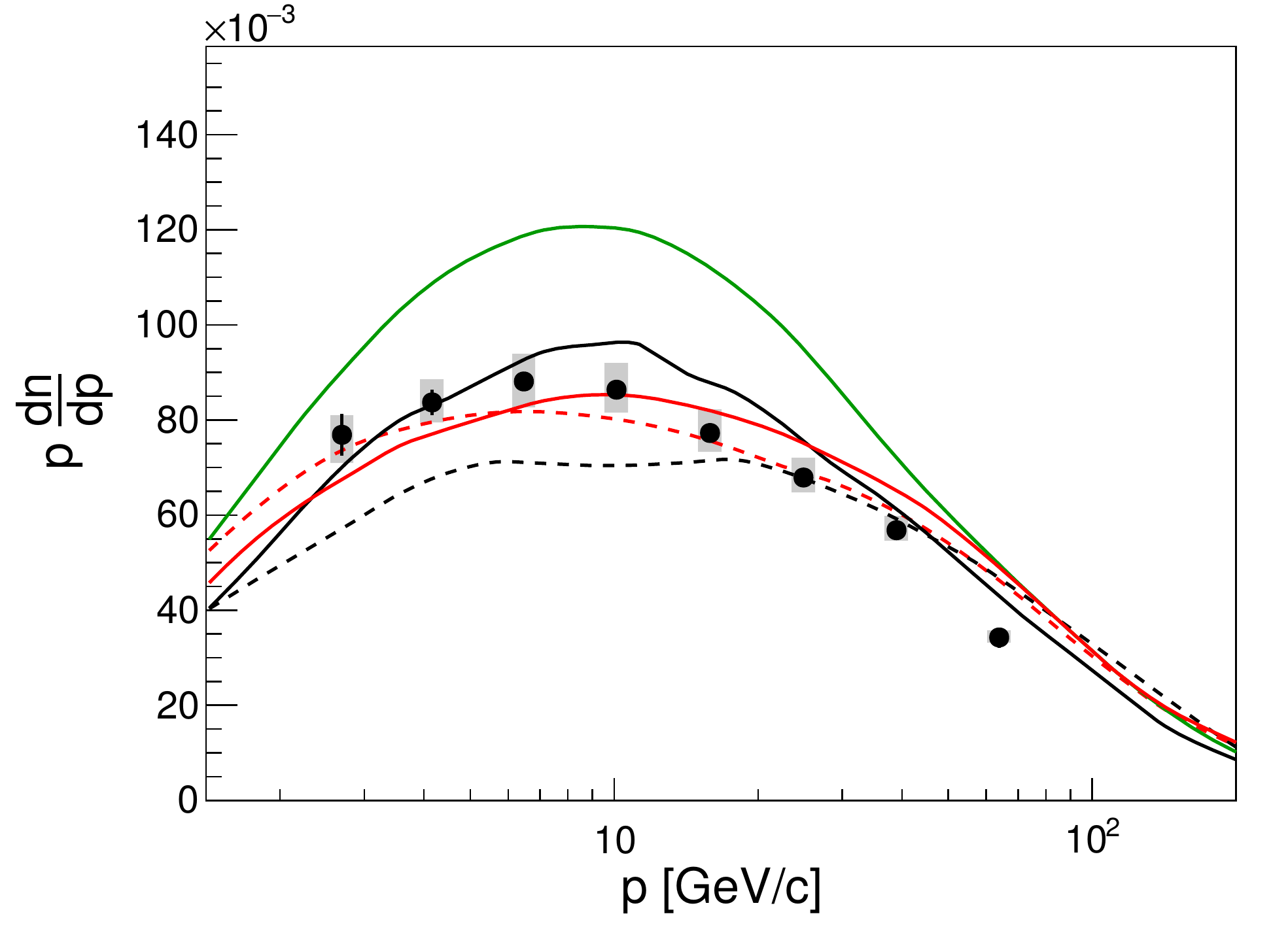}
    \put(28,63){$\pi^-$+C$\rightarrow$\kzeros+X}
    \put(29,58){at 350 \GeVc}
  \end{overpic}

  \caption{Spectra of \kzeros as a function of \pp (integrated over \pT),
    for the 158 and 350 \GeVc data set.
    The statistical uncertainties are shown as black bars and the systematic ones as gray bands.}
  \label{fig:vzero:kzeros}
\end{figure*}

\begin{figure*}
  \centering
  \begin{overpic}[clip, rviewport=0 0 1 1,width=0.47\textwidth]{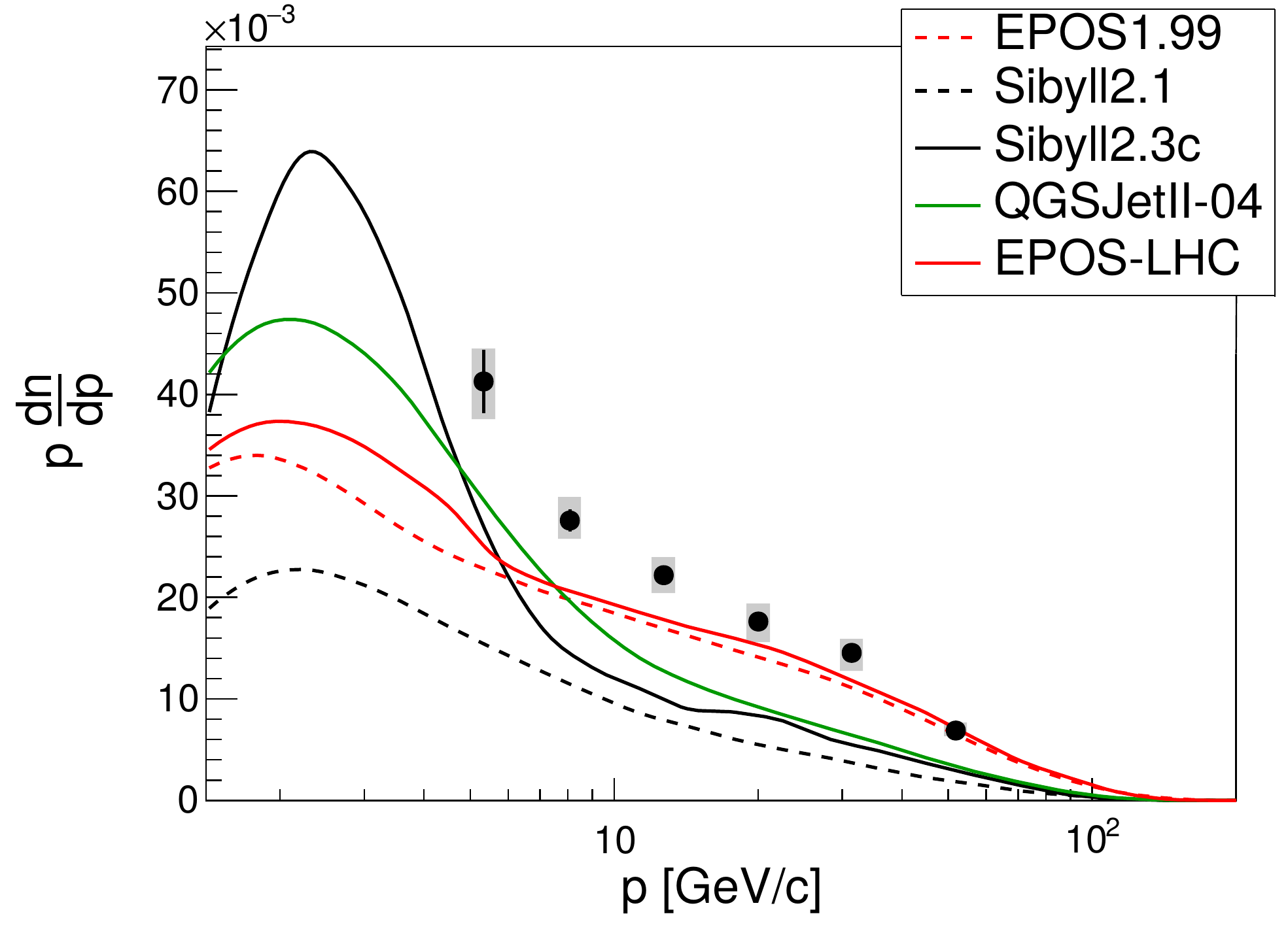}
    \put(70,75){\Large\NASixtyOne PRELIMINARY}
    \put(36,63){$\pi^-$+C$\rightarrow$\lamb+X}
    \put(37,58){at 158 \GeVc}
  \end{overpic}
  \begin{overpic}[clip, rviewport=0 0 1 1,width=0.47\textwidth]{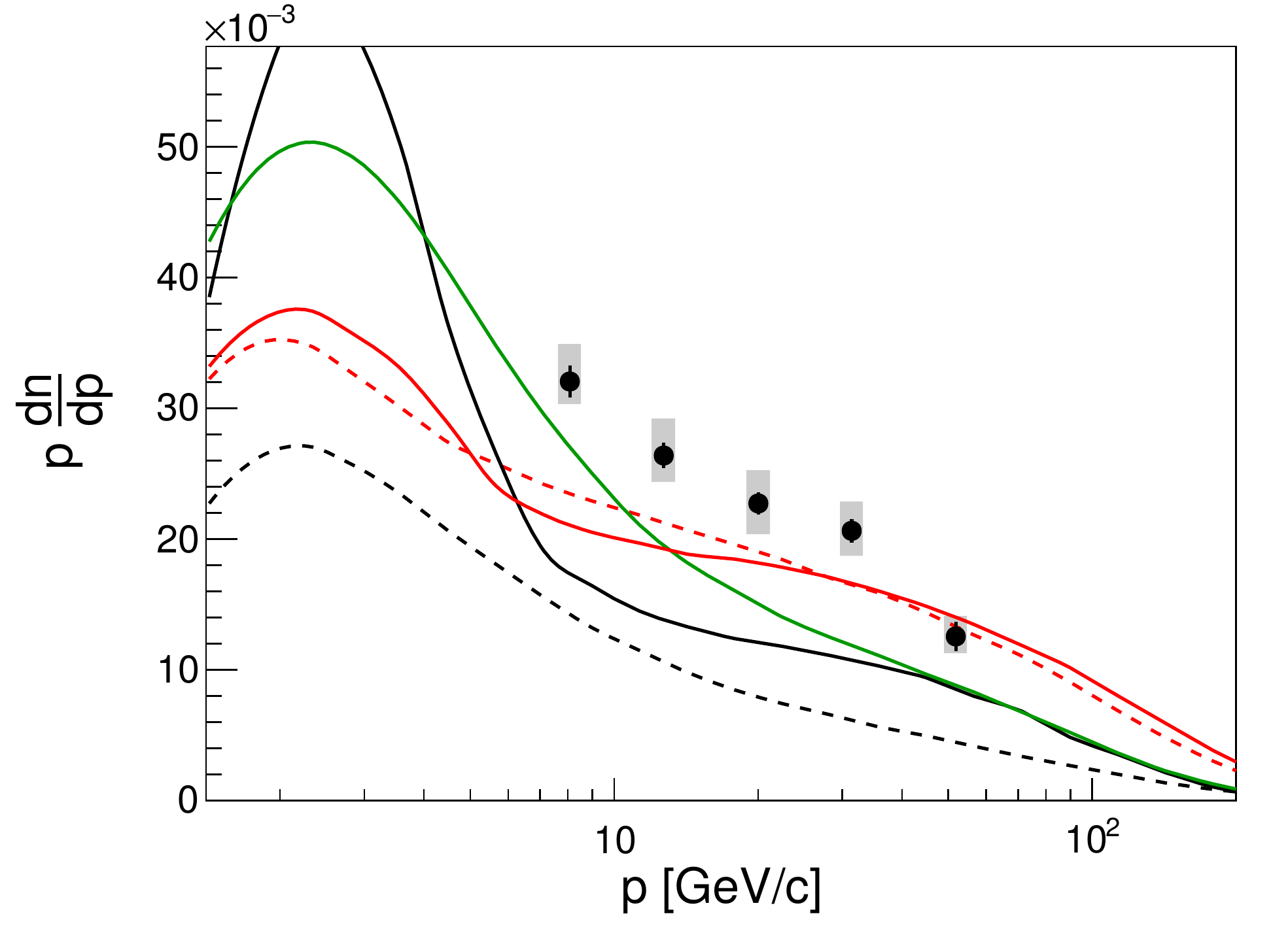}
    \put(36,63){$\pi^-$+C$\rightarrow$\lamb+X}
    \put(37,58){at 350 \GeVc}
  \end{overpic}

  \begin{overpic}[clip, rviewport=0 0 1 1,width=0.47\textwidth]{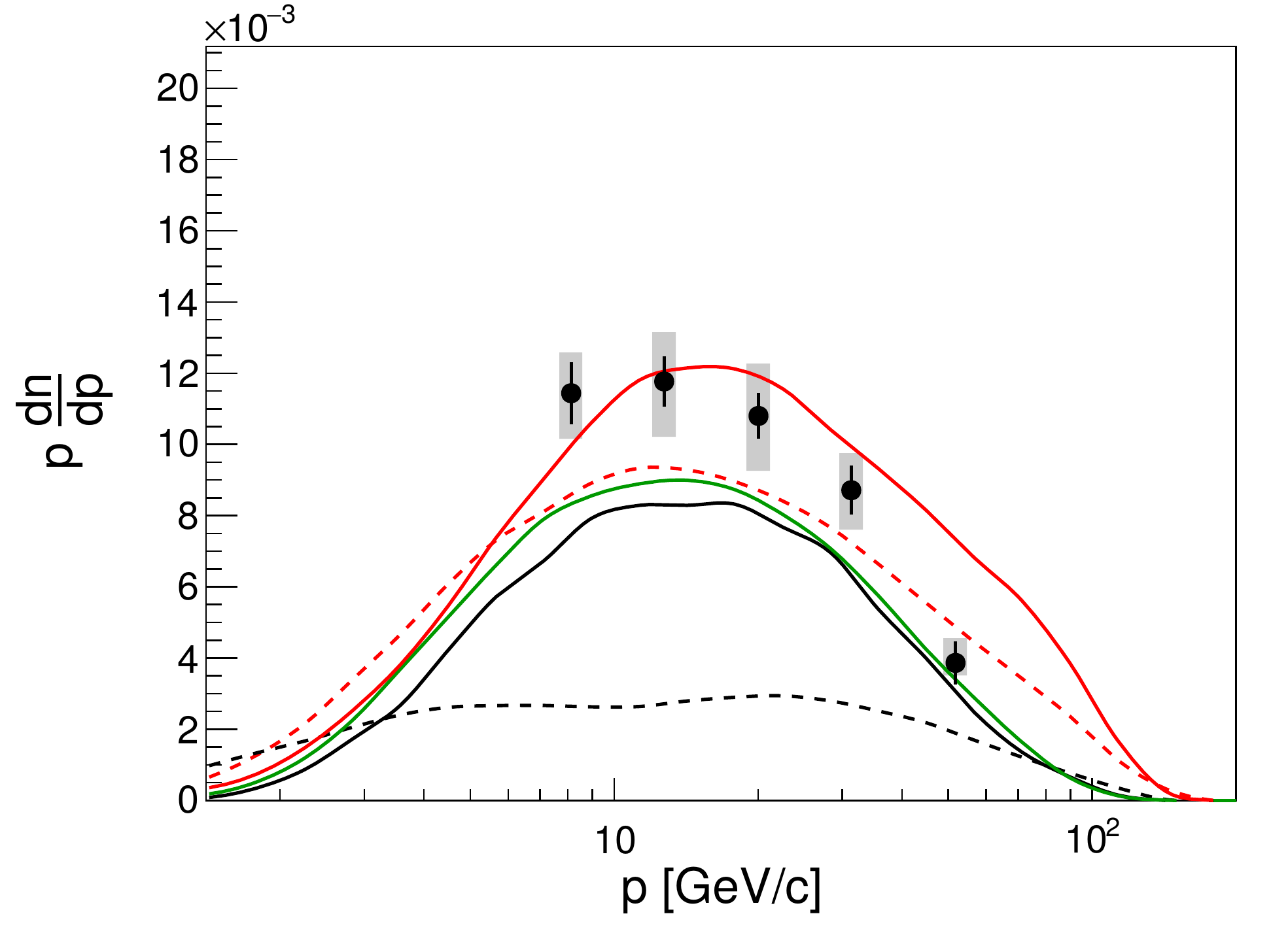}
    \put(36,63){$\pi^-$+C$\rightarrow$\antilamb+X}
    \put(37,58){at 158 \GeVc}
  \end{overpic}
  \begin{overpic}[clip, rviewport=0 0 1 1,width=0.47\textwidth]{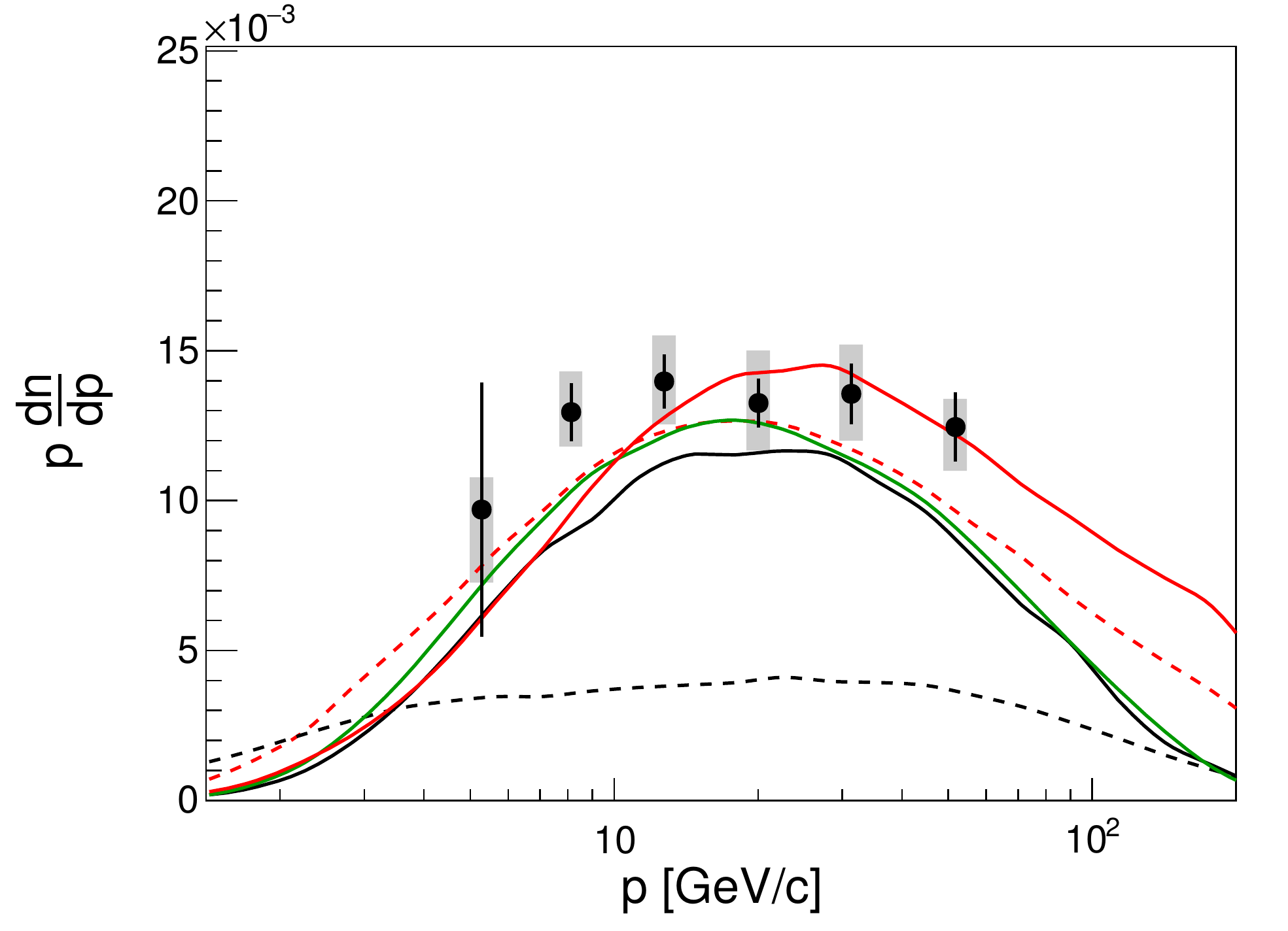}
    \put(36,63){$\pi^-$+C$\rightarrow$\antilamb+X}
    \put(37,58){at 350 \GeVc}
  \end{overpic}

  \caption{Spectra of \lambs as a function of \pp (integrated over \pT),
    for the 158 and 350 \GeVc data set.
    The statistical uncertainties are shown as black bars and the systematic ones as gray bands.}
  \label{fig:vzero:lamb}
\end{figure*}

This analysis was performed in 2-dimensional
phase space bins of \pp and \pT. For each phase space bin,
the detector effects were corrected by using Monte Carlo simulations.
The full double-differential spectra
as a function of \pp and \pT for \lambs and \kzeros
at 158 and 350 \GeVc can be found in Ref.~\cite{PradoHEP2018}. 
In~\cref{fig:vzero:lamb,fig:vzero:kzeros} we show the measured 
single-differential spectra as function of \pp (integrated over \pT)
together with predictions of the hadronic models.

%%%%%%%%%%%%%%%%%%%%%%%%%%%%%%%%
\section{Production of \r0, \kstar and $\omega$}
\label{sec:resonance}

\begin{figure}[!h]
  \centering

  \begin{overpic}[clip, rviewport=0 0 1 1,width=0.495\textwidth]{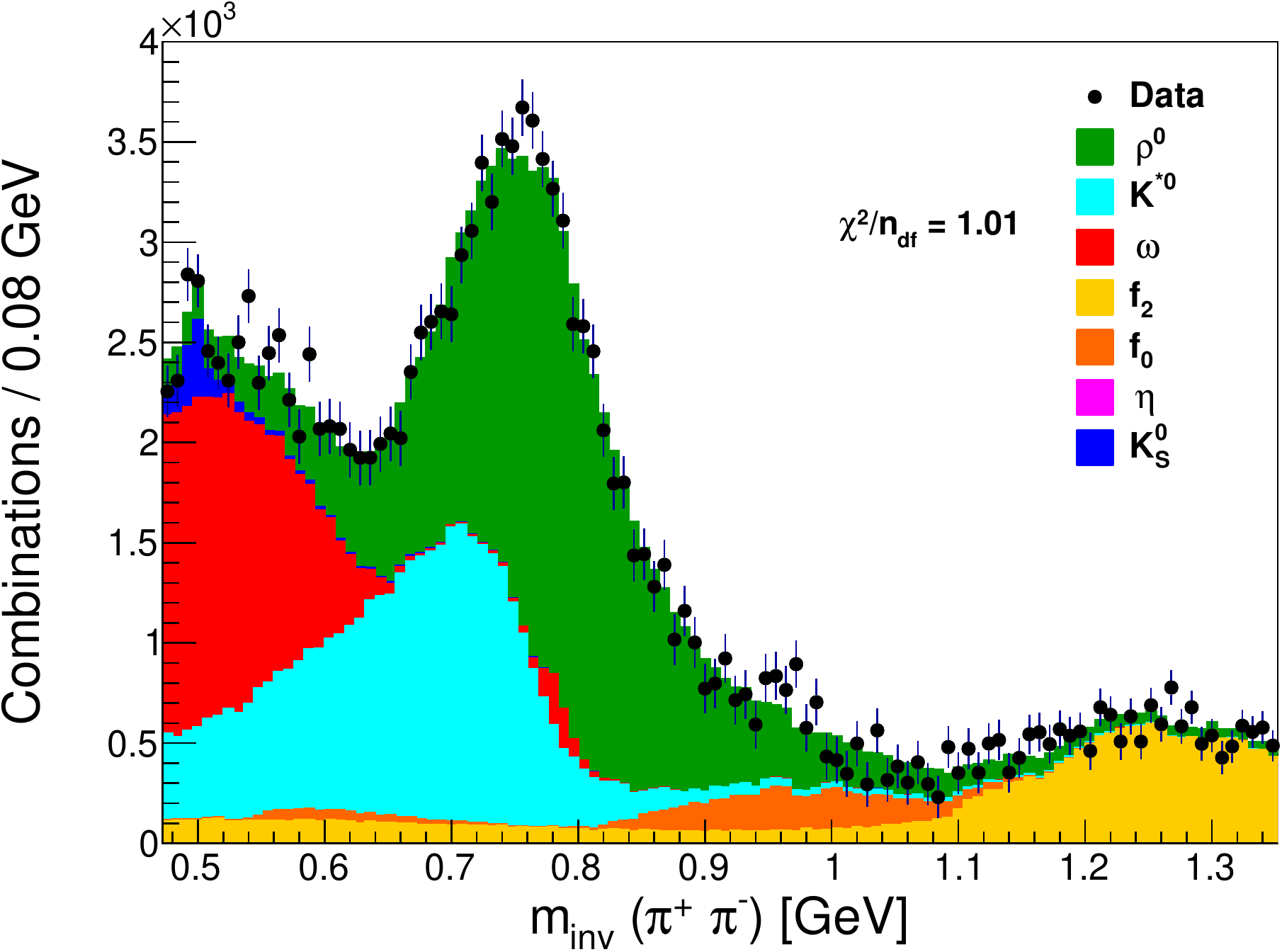}
    \put(17,82){}
  \end{overpic}

  \caption{Example of the \minvpi distribution for one
    \xF bin ($0.3 < \xF < 0.4$) of the 158 \GeVc data set.
    The black markers show the measured distributions and the colored
    distributions show the results of the template fit.}
  \label{fig:resonance:mass}
\end{figure}

By using the \NASixtyOne apparatus, the yields of \r0, \kstar and $\omega$
can be measured through the \pipi invariant mass (\minvpi) spectra. 
The signal extraction is performed by fitting Monte Carlo templates
to the measured \minvpi distribution. 
The Monte Carlo events were generated using \EposLong as hadronic
interaction model and they were passed through the full \NASixtyOne
detector simulation and reconstruction chain.
The estimation of the combinatorial background were done
by two methods: the charge mixing method, in which the
$\pi^+\pi^+$ and $\pi^-\pi^-$ are
treated as the background, and the Monte Carlo method, in which the
background mass distribution is obtained directly from
simulations. One example of the \minvpi distributions with
the results of the template fit is shown in~\cref{fig:resonance:mass}.
After the signal extraction, the particle yields
were corrected by the detector effects and the production
spectra were derived. The full description of the analysis procedure
and the results can be found in Ref.~\cite{\RhoPaper}.

We show in~\cref{fig:resonance:spec} the obtained \r0, \kstar and $\omega$ spectra
together with predictions from simulations with the hadronic
models.
The \r0 spectra are shown for both beam energies,
158 and 350 \GeVc, and the $\omega$ and \kstar spectra are limited to the 158 \GeVc
data set because of the large uncertainties obtained at 350 \GeVc.

%%%%%%%%%%%%%%%%%%%%%%%%%%%%%%%%
\section{Summary and conclusions}
\label{sec:summary}

The \NASixtyOne experiment, within its very rich program,
has provided a large number of measurements which
have been used for testing and tuning of hadronic interaction
models used by the cosmic ray community. 
In this paper, we have summarized the results of the
special cosmic ray runs for \piC interactions.

\begin{figure*}
  \centering
  \begin{overpic}[clip, rviewport=0.01 0.125 0.97 0.92,width=0.495\textwidth]{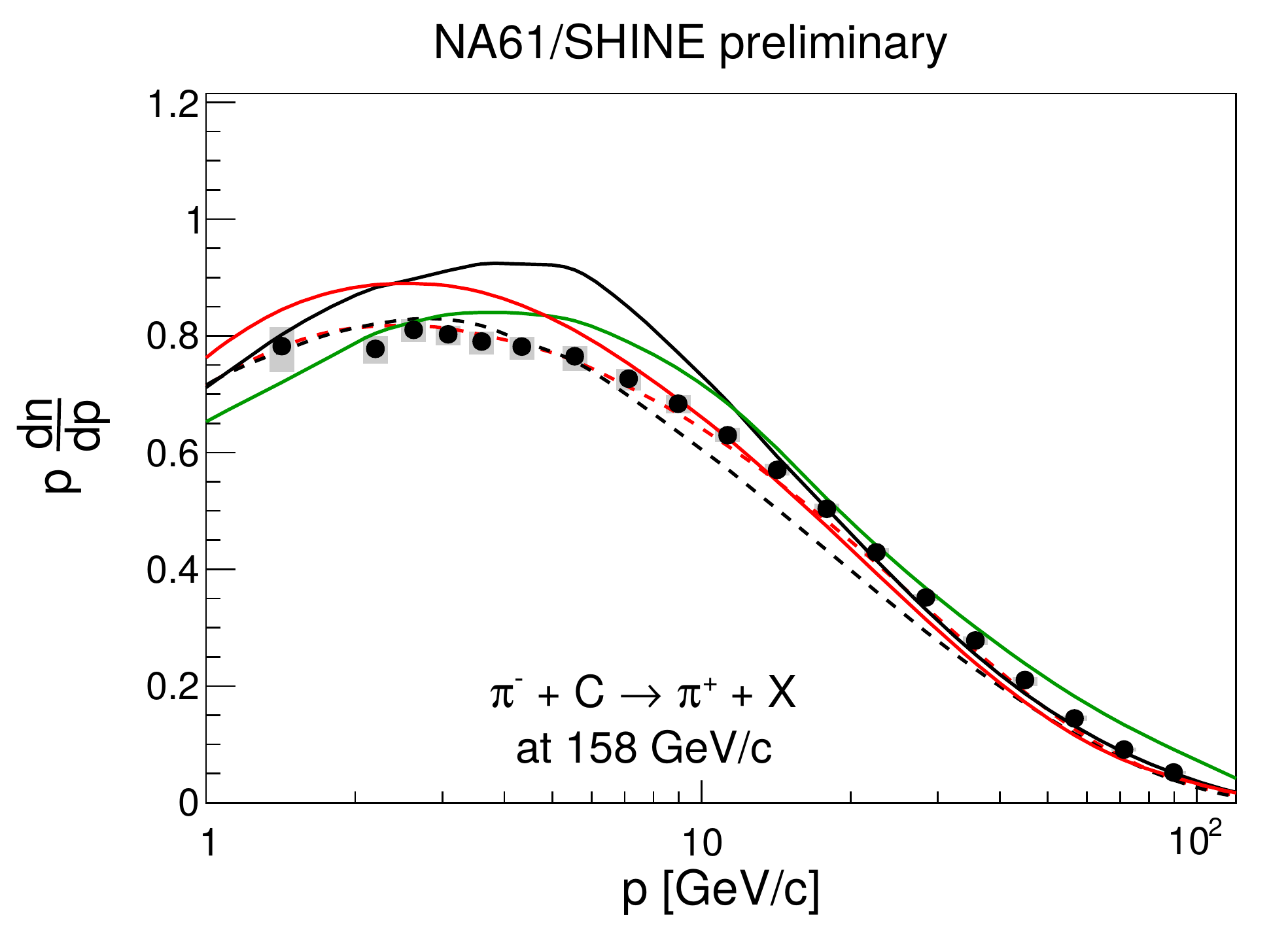}
    \put(58,50){\Large\NASixtyOne}
    \put(56,43){\Large PRELIMINARY}
  \end{overpic}
  \begin{overpic}[clip, rviewport=0.01 0.125 0.97 0.92,width=0.495\textwidth]{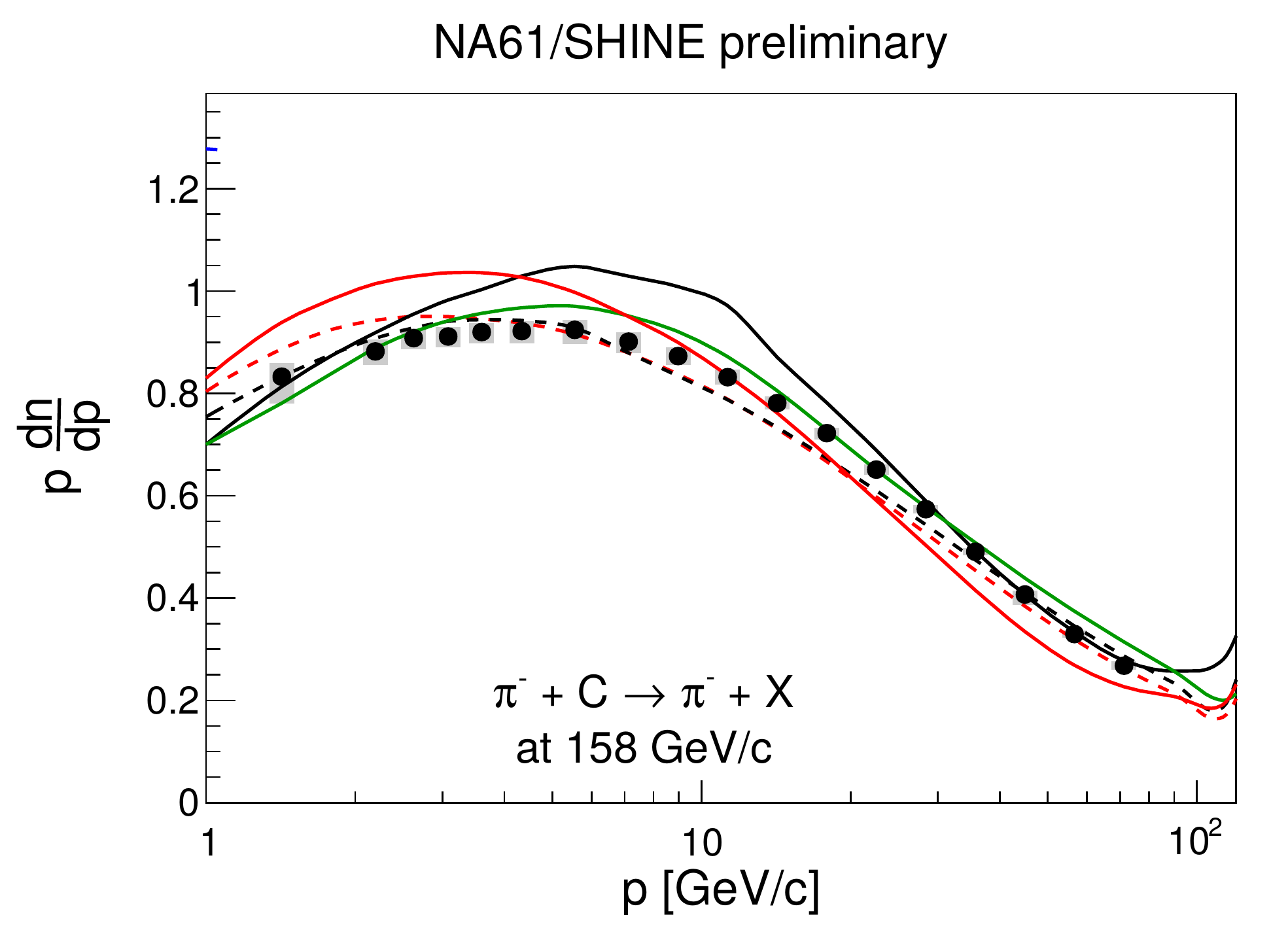}
    \put(15.5,23){}
  \end{overpic}

  \begin{overpic}[clip, rviewport=0.01 0.125 0.97 0.91,width=0.495\textwidth]{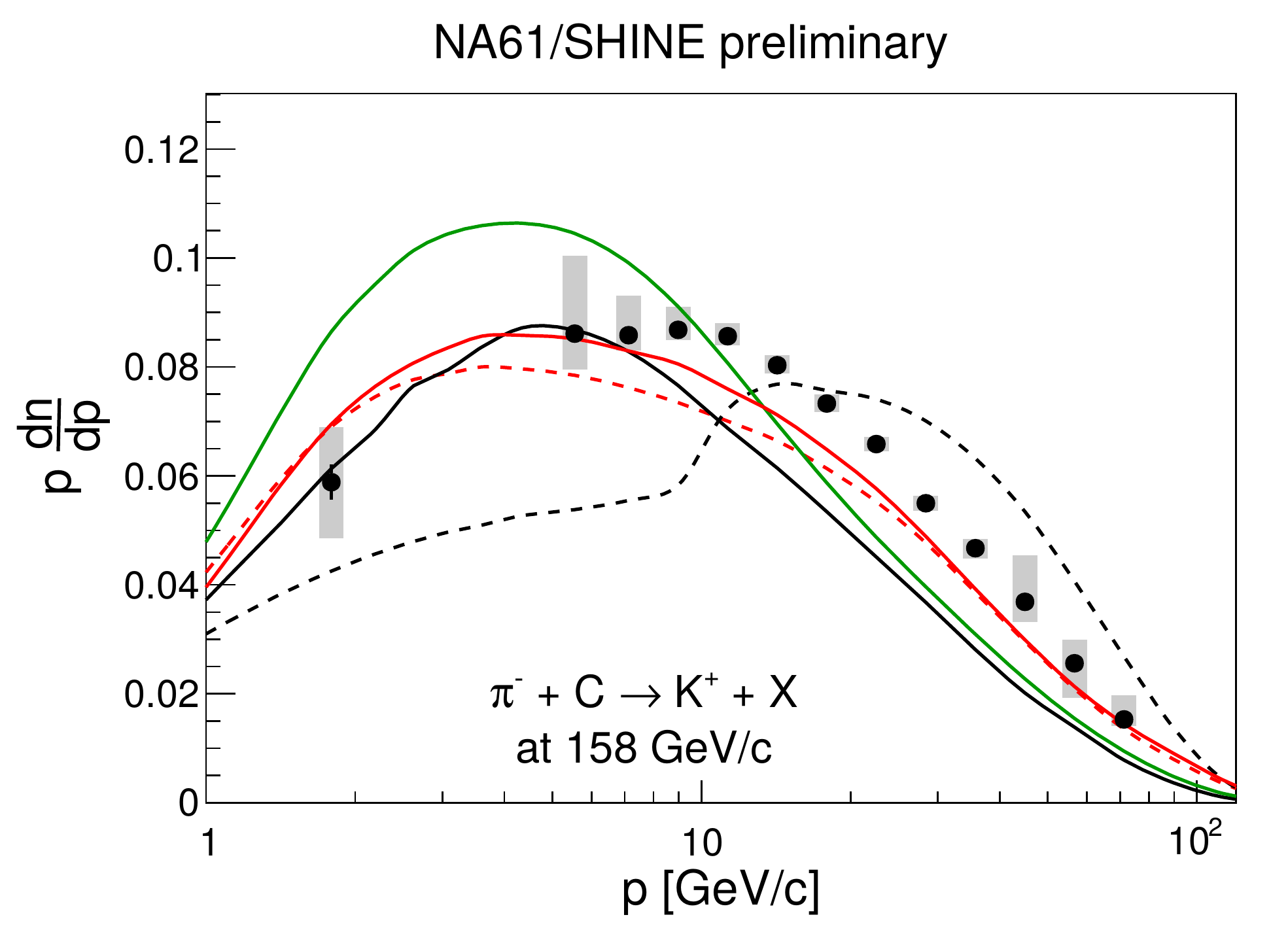}
    \put(15.5,23){}
  \end{overpic}
  \begin{overpic}[clip, rviewport=0.01 0.125 0.97 0.91,width=0.495\textwidth]{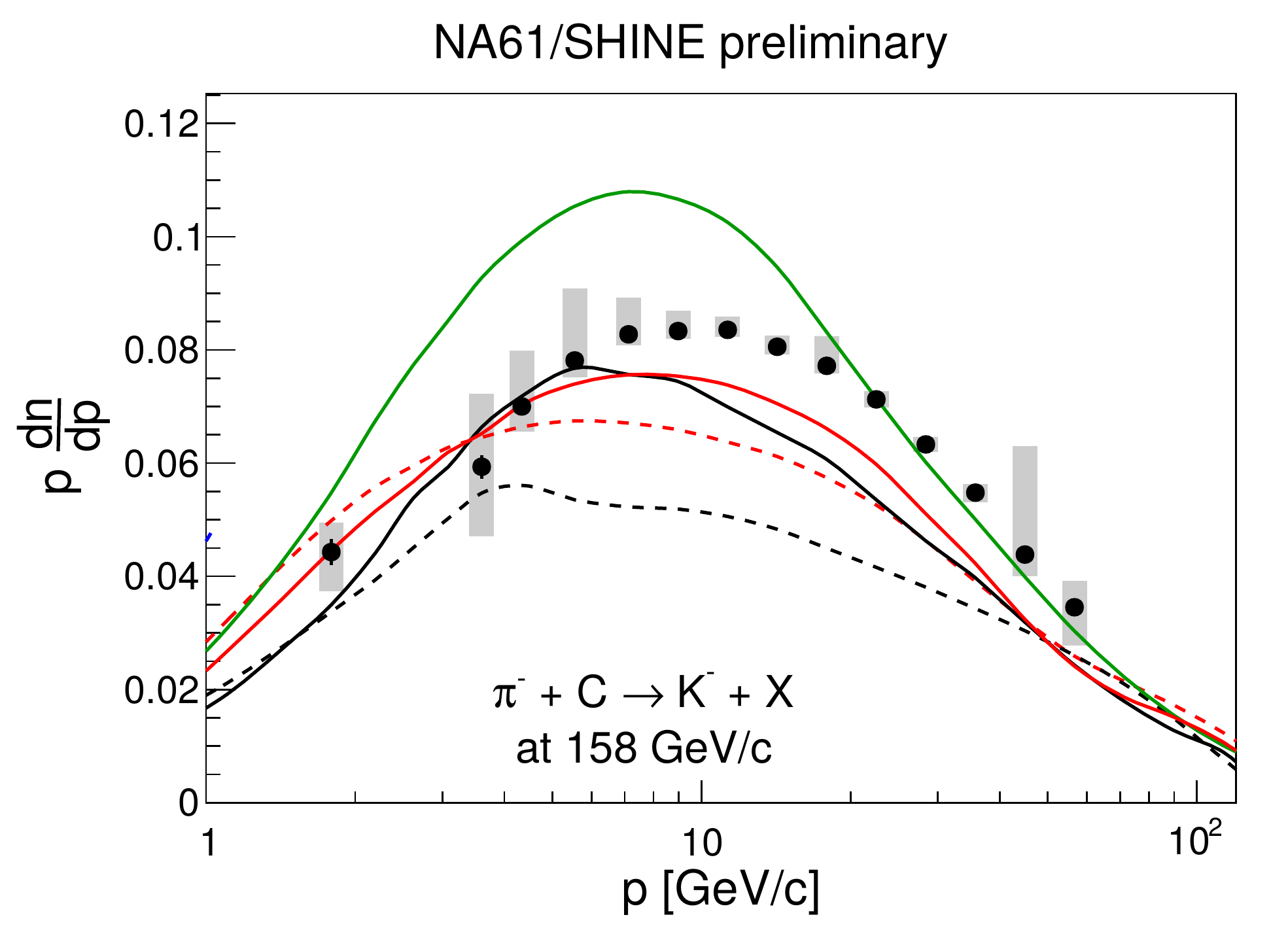}
    \put(15.5,23){}
  \end{overpic}

  \begin{overpic}[clip, rviewport=0.01 0 0.97 0.91,width=0.495\textwidth]{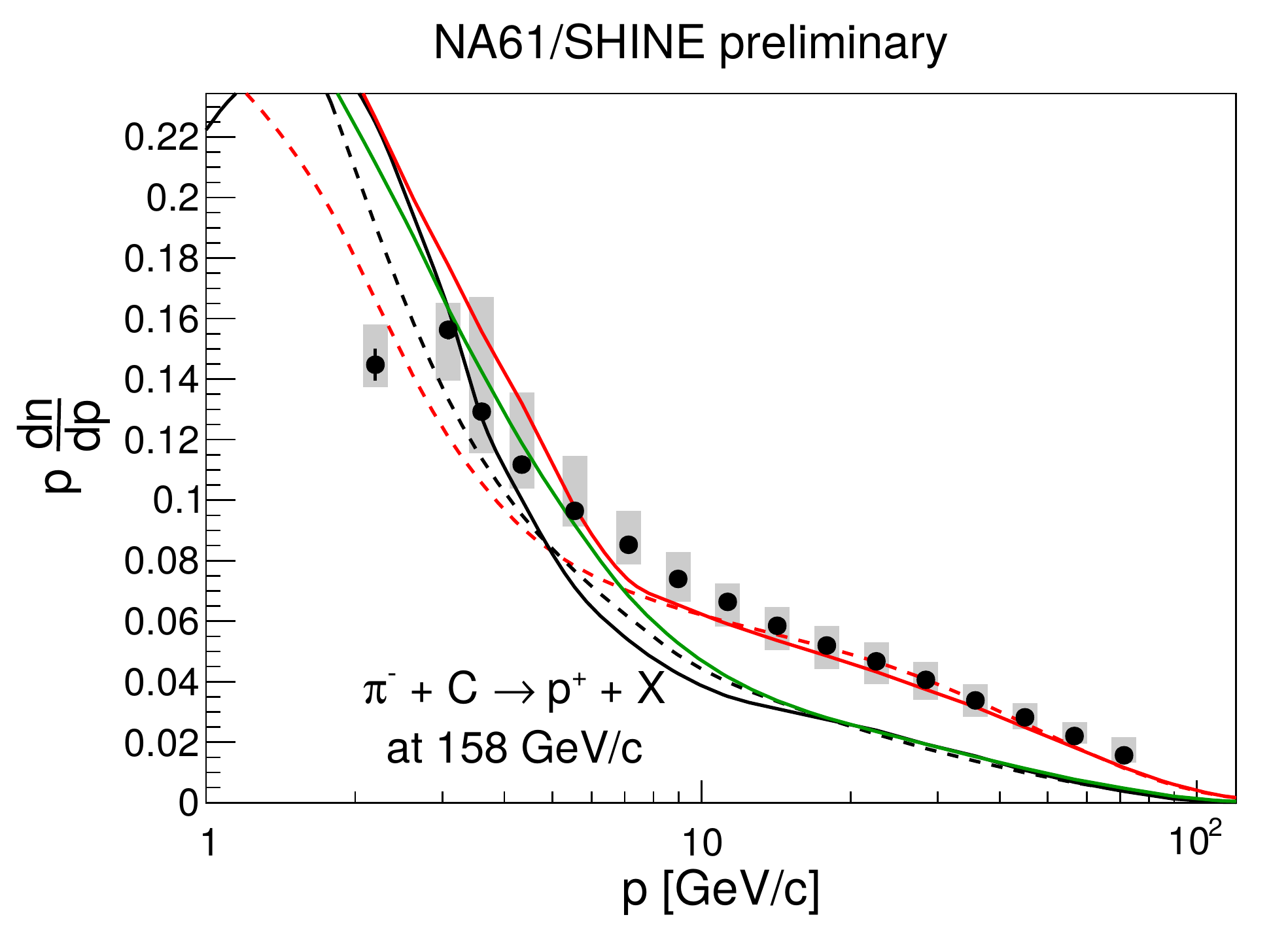}
    \put(60,30){
      \includegraphics[clip, rviewport=0.65 0.5 1 1, width=0.16\textwidth]{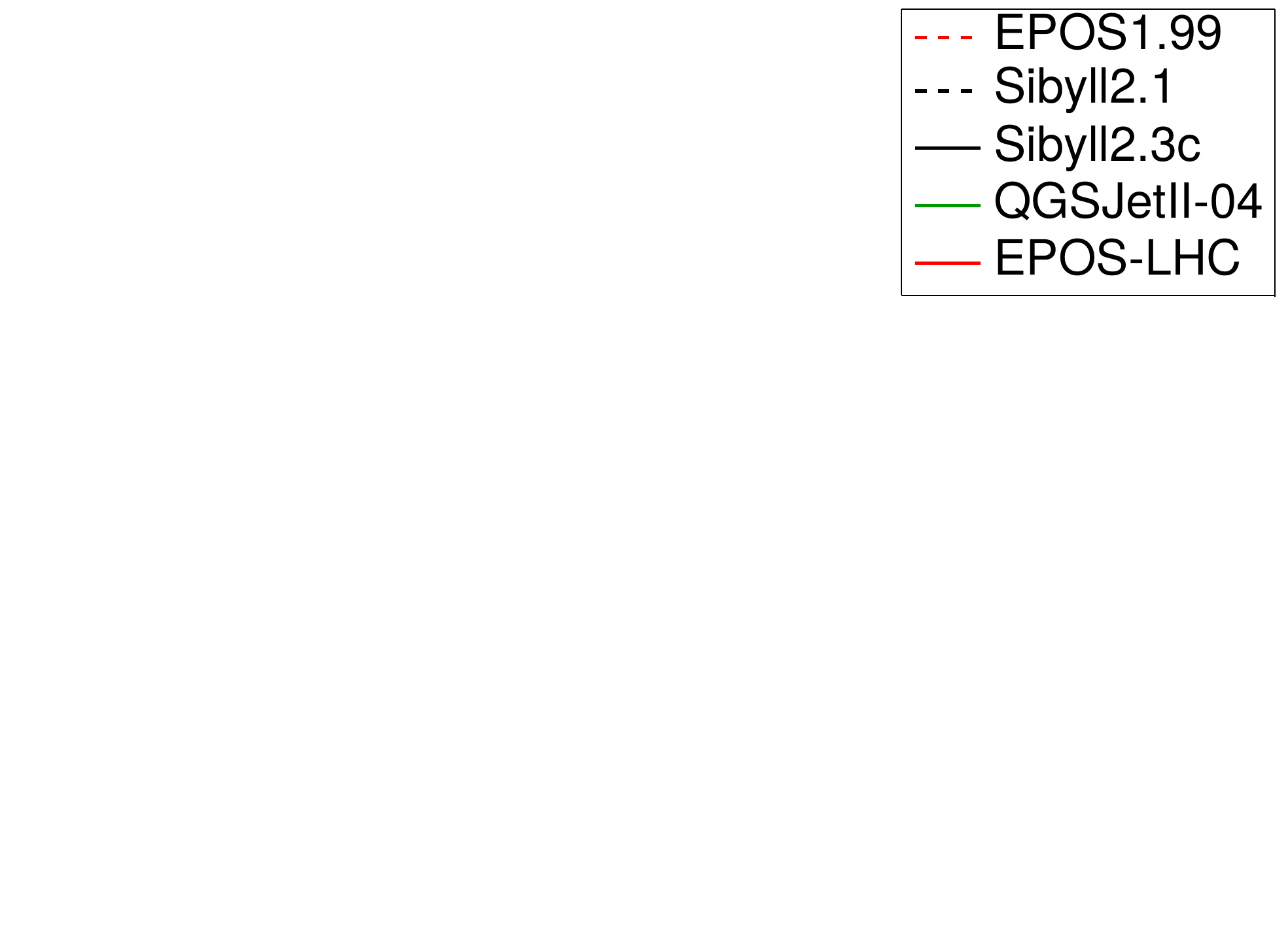}
    }
  \end{overpic}
  \begin{overpic}[clip, rviewport=0.01 0 0.97 0.91,width=0.495\textwidth]{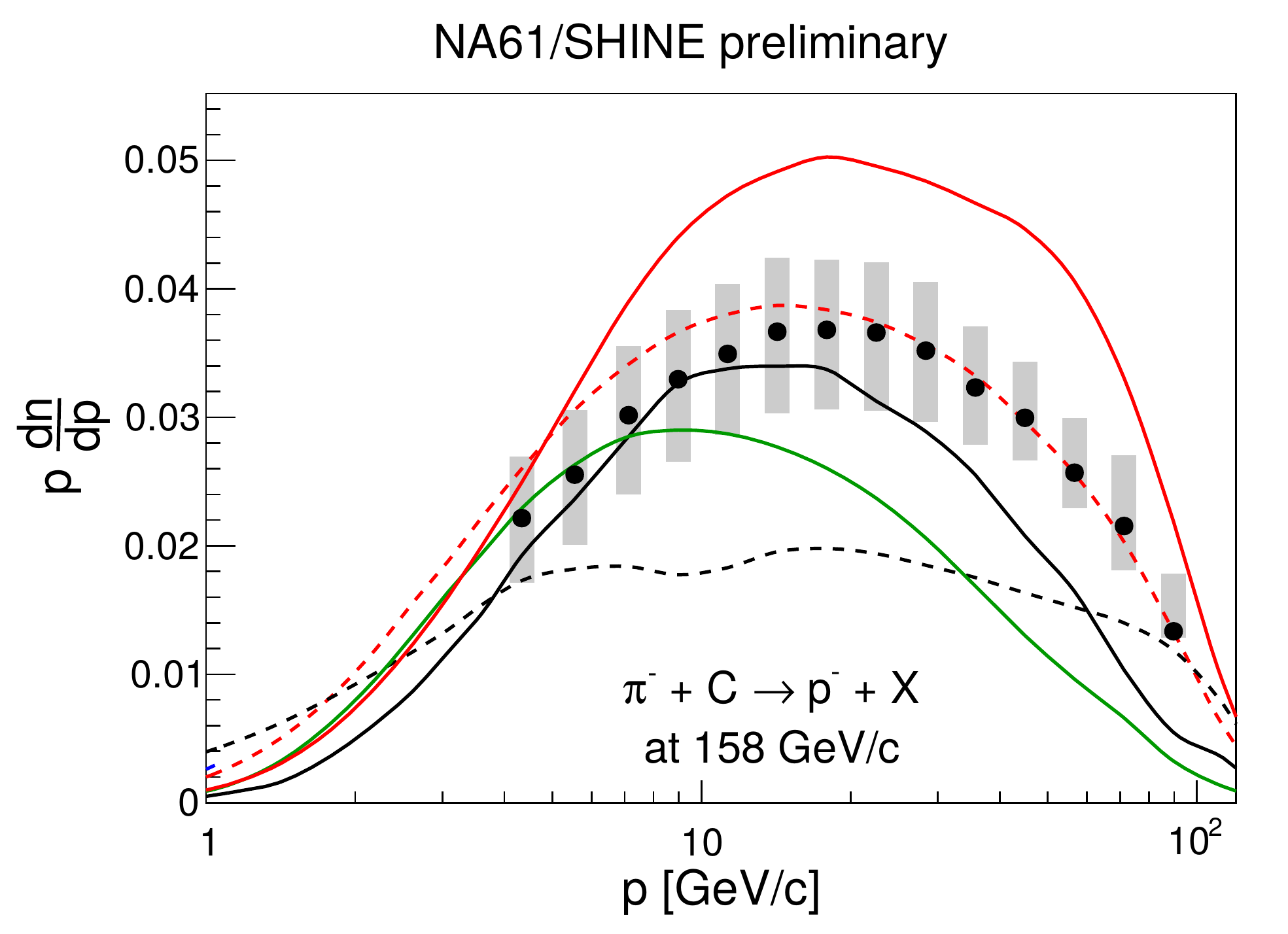}
    \put(15.5,23){}
  \end{overpic}

  \caption{Spectra of \pions, \kaons and \protons as a function of \pp (integrated over \pT),
    for the 158 \GeVc data set.
    The statistical uncertainties are shown as black bars and the systematic ones as gray bands.}
  \label{fig:hadron:int158}
\end{figure*}

\begin{figure*}
  \centering
  \begin{overpic}[clip, rviewport=0.01 0.125 0.97 0.92,width=0.495\textwidth]{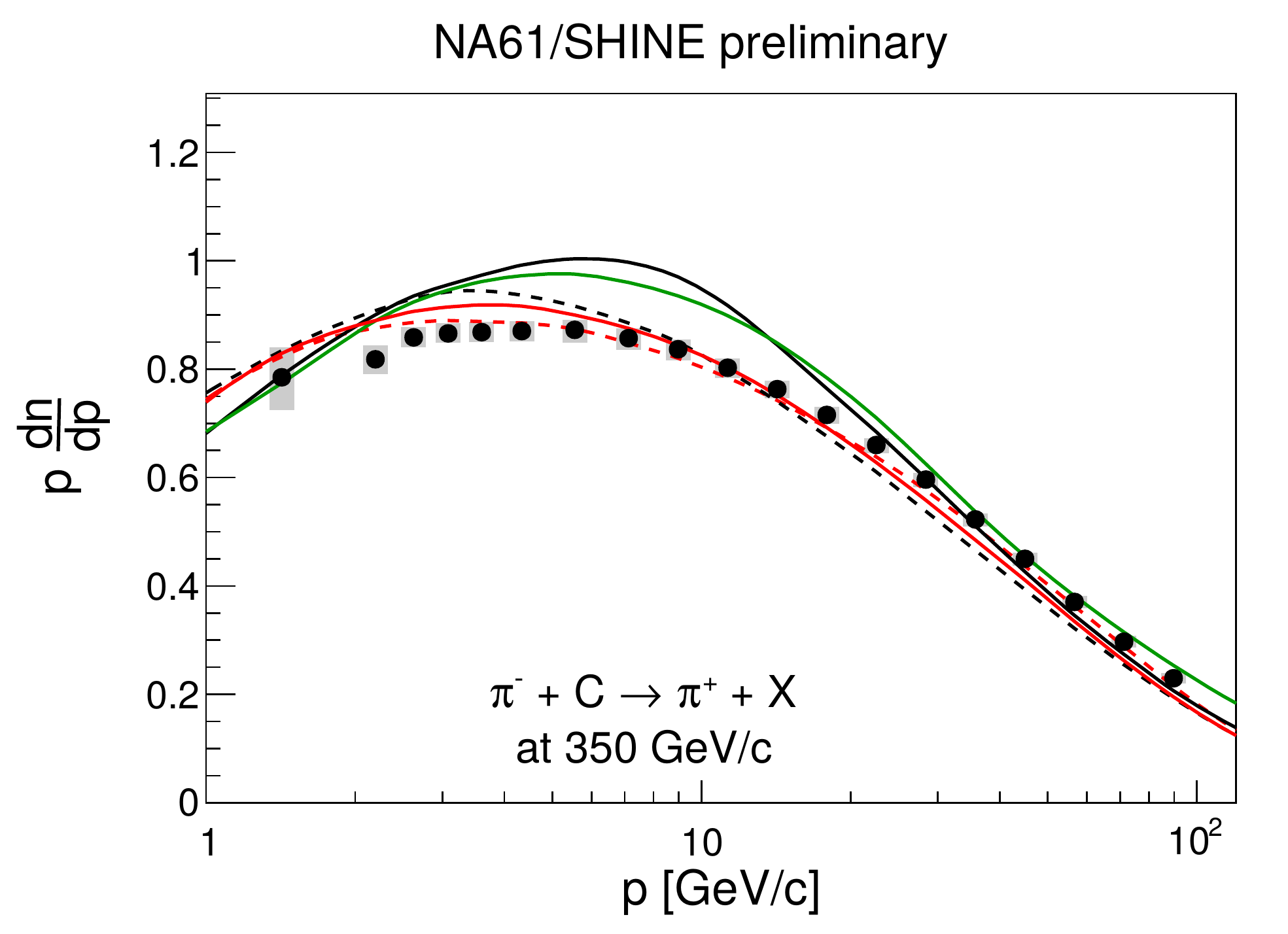}
    \put(58,50){\Large\NASixtyOne}
    \put(56,43){\Large PRELIMINARY}
  \end{overpic}
  \begin{overpic}[clip, rviewport=0.01 0.125 0.97 0.92,width=0.495\textwidth]{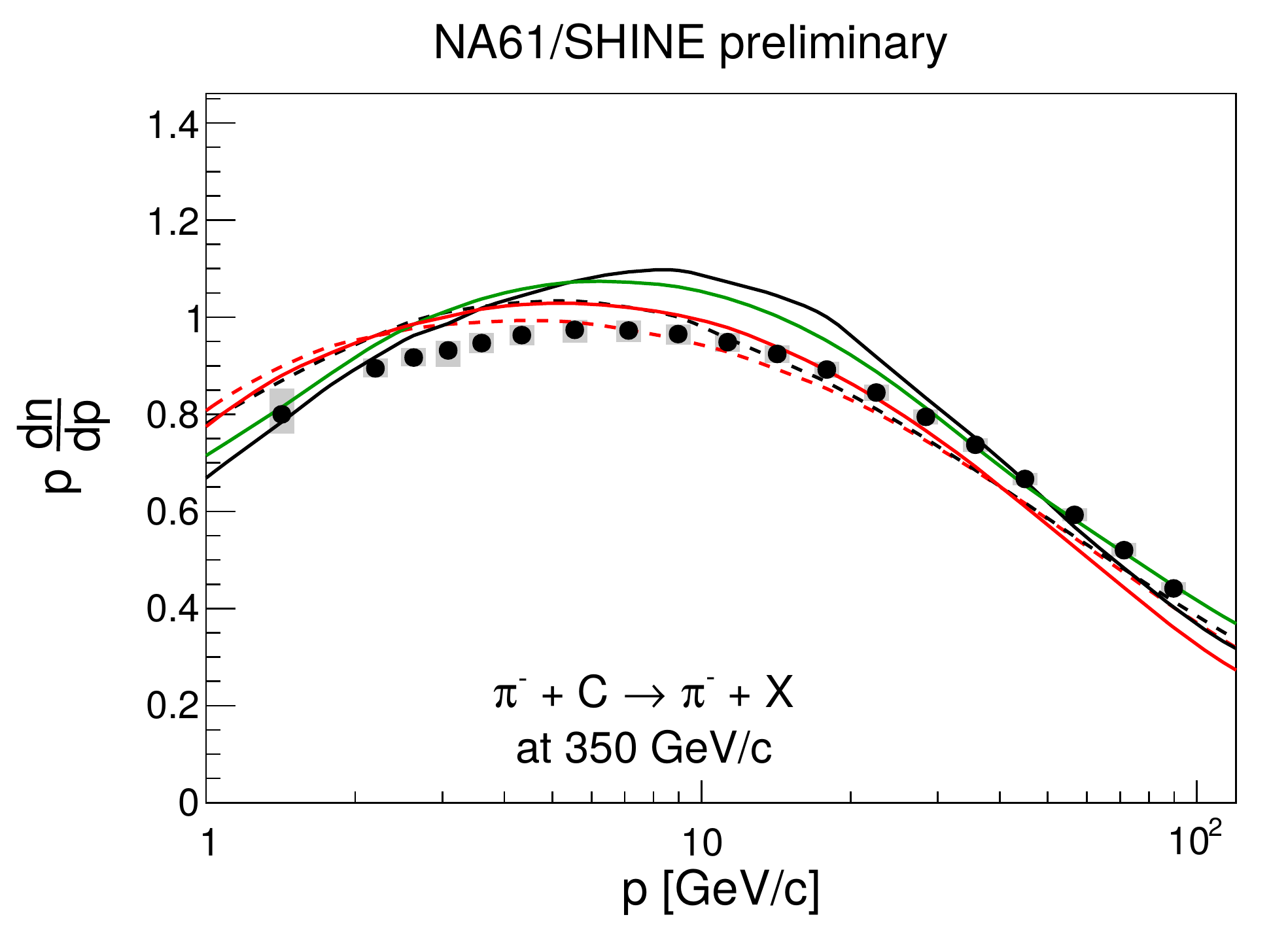}
    \put(15.5,23){}
  \end{overpic}

  \begin{overpic}[clip, rviewport=0.01 0.125 0.97 0.91,width=0.495\textwidth]{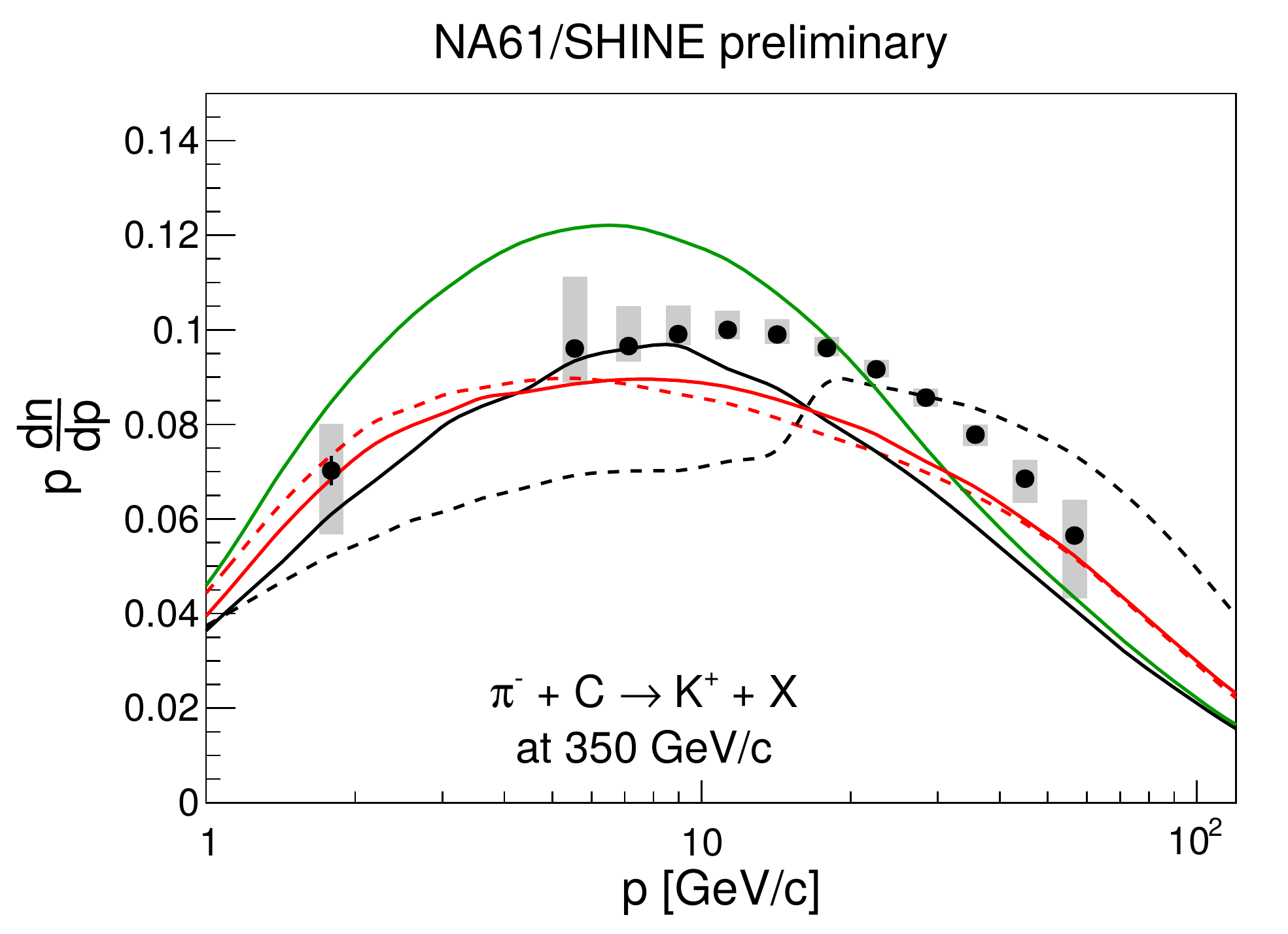}
    \put(15.5,23){}
  \end{overpic}
  \begin{overpic}[clip, rviewport=0.01 0.125 0.97 0.91,width=0.495\textwidth]{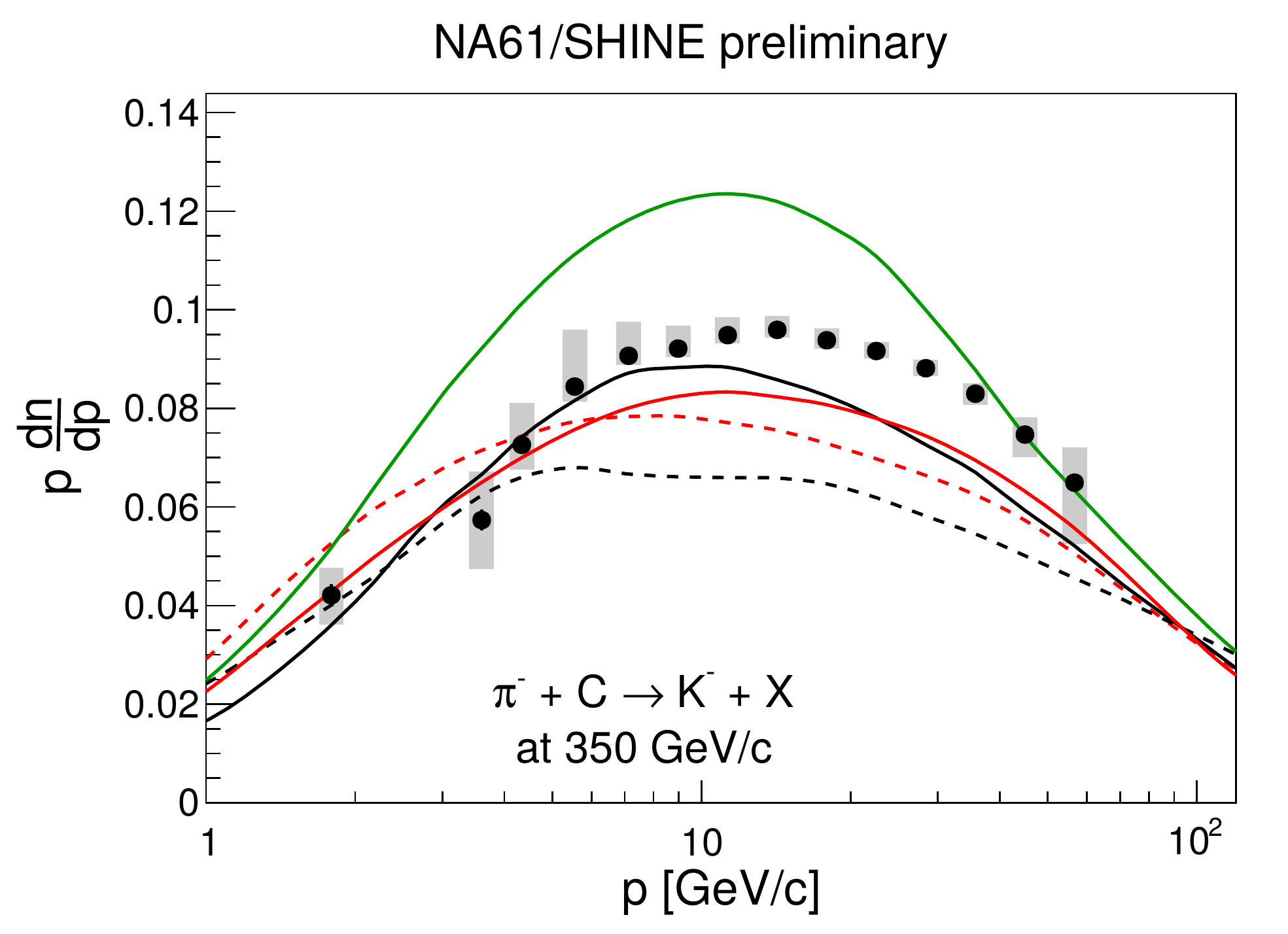}
    \put(15.5,23){}
  \end{overpic}

  \begin{overpic}[clip, rviewport=0.01 0 0.97 0.91,width=0.495\textwidth]{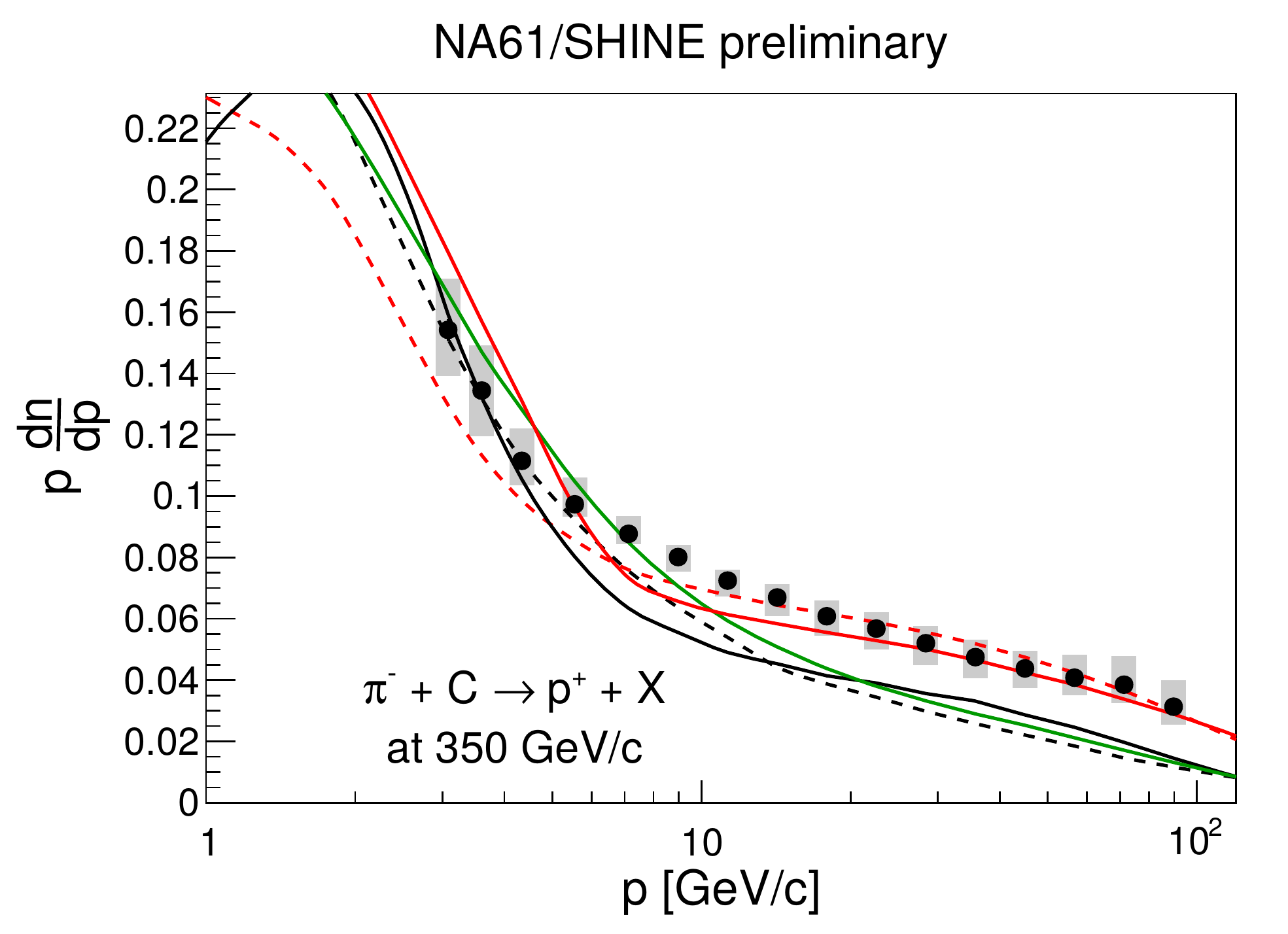}
    \put(60,30){
      \includegraphics[clip, rviewport=0.65 0.5 1 1, width=0.16\textwidth]{figures/new_leg}
    }
  \end{overpic} 
  \begin{overpic}[clip, rviewport=0.01 0 0.97 0.91,width=0.495\textwidth]{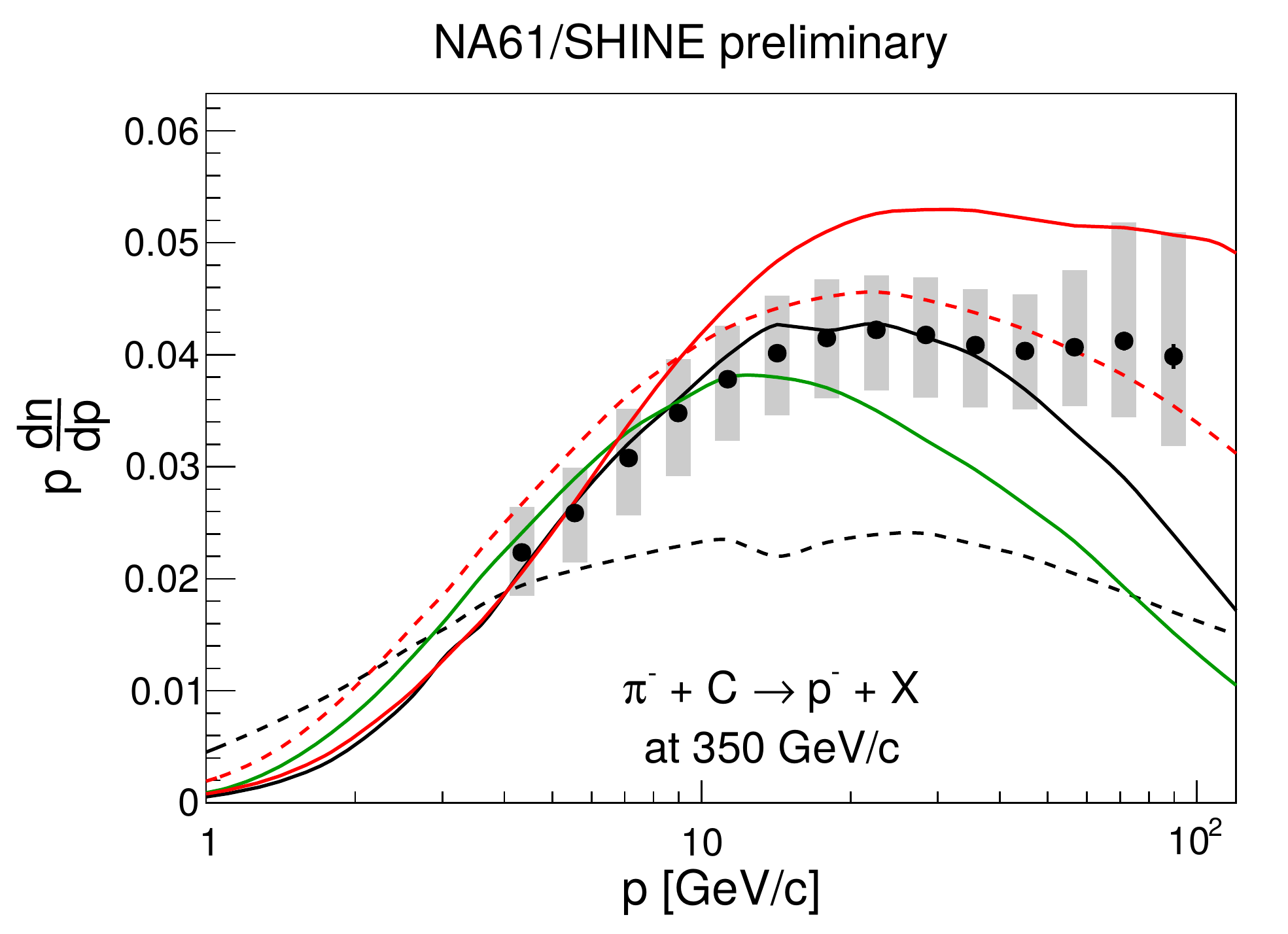}
    \put(15.5,23){}
  \end{overpic}

  \caption{Spectra of \pions, \kaons and \protons as a function of \pp (integrated over \pT),
    for the 350 \GeVc data set.
    The statistical uncertainties are shown as black bars and the systematic ones as gray bands.}
  \label{fig:hadron:int350}
\end{figure*}

First, we have shown the identified spectra of charged hadrons 
obtained by using the \dedx measurements. Of particular interest here
is the production spectra of \protons, which are relevant
to study the (anti)baryon productions in hadron-air interactions
and its implications on the muon production in EAS.
From the \antiproton spectra shown
in~\cref{fig:hadron:int158,fig:hadron:int350},
one can see that the (anti)baryon production is not underestimated
in general by the models. In particular, the \Epos model shows to describe
very well the \antiproton production. As a conclusion, the 
underproduction of (anti)baryons in $\pi$-air interactions
by the hadronic models
is unlikely to be the most relevant source of the lack of muons 
in simulations.  

Secondly, we have shown the results of the \vzero analysis, aiming
the \lambs and \kzeros spectra. Although these measurements are
surely relevant for model testing and tuning, our main motivation here
is to reduce the systematic uncertainties on the \pions and \protons spectra
due to the feed-down contributions from weak decays.
Since a significant fraction of \pions and \protons detected
are produced by the decay of \lamb, \antilamb and \kzeros, this effect has to be
corrected. In the results shown in~\cref{sec:hadrons} (and in Ref.\cite{\RaulICRC})
this correction is done by using Monte Carlo simulations and the model
dependence of this procedure is added to the systematic uncertainties.
By measuring the spectra of \lamb, \antilamb and \kzeros, we are able
to avoid this model dependence and consequently reduce the systematic uncertainties.
Updated \pions and \protons spectra with improved
systematic uncertainties will be presented
in the future in another publication.

\begin{figure*}
  \centering
  \begin{overpic}[clip, rviewport=0 0 1 1,width=0.47\textwidth]{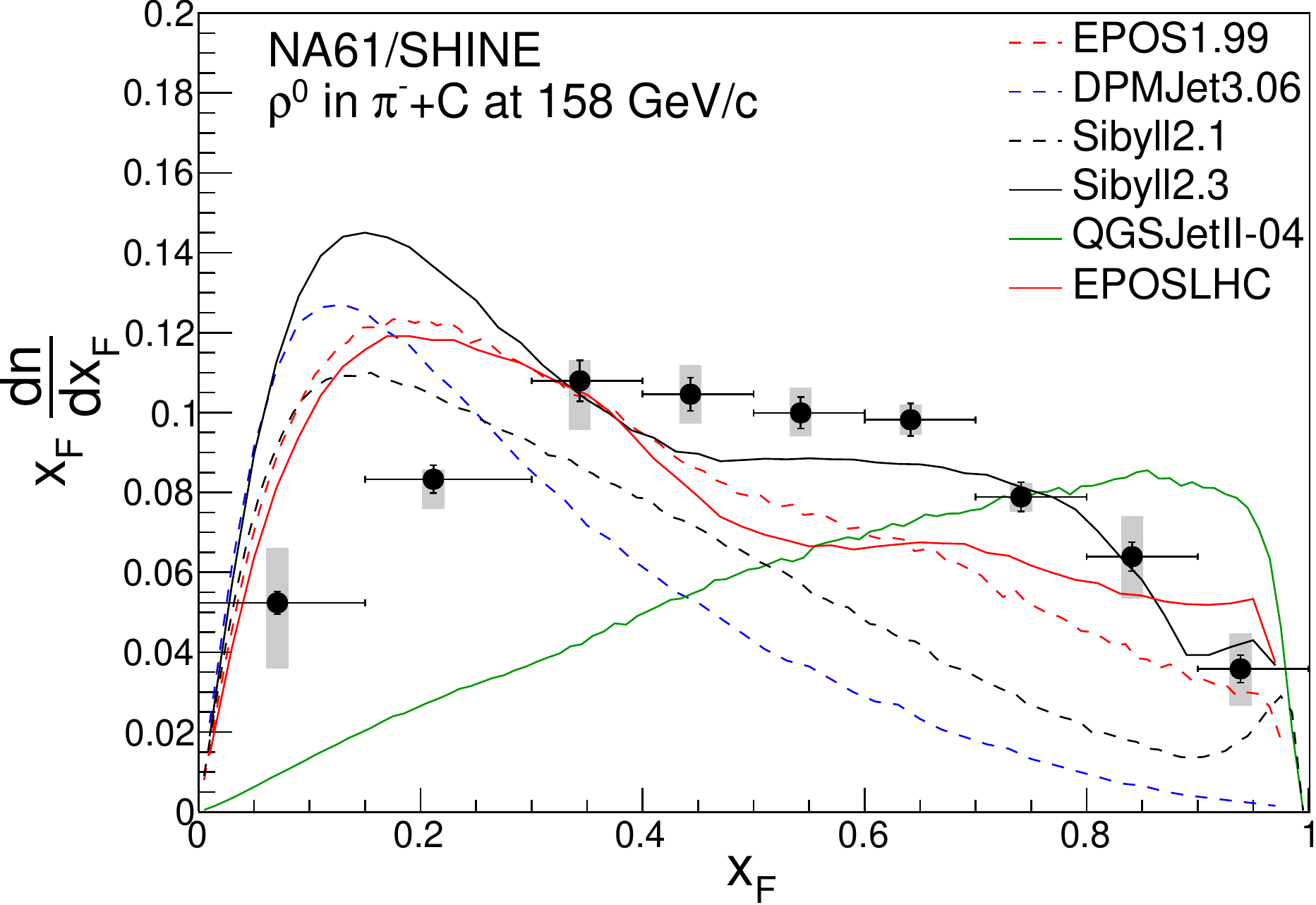}
    \put(58,50){}
  \end{overpic}
  \begin{overpic}[clip, rviewport=0 0 1 1,width=0.47\textwidth]{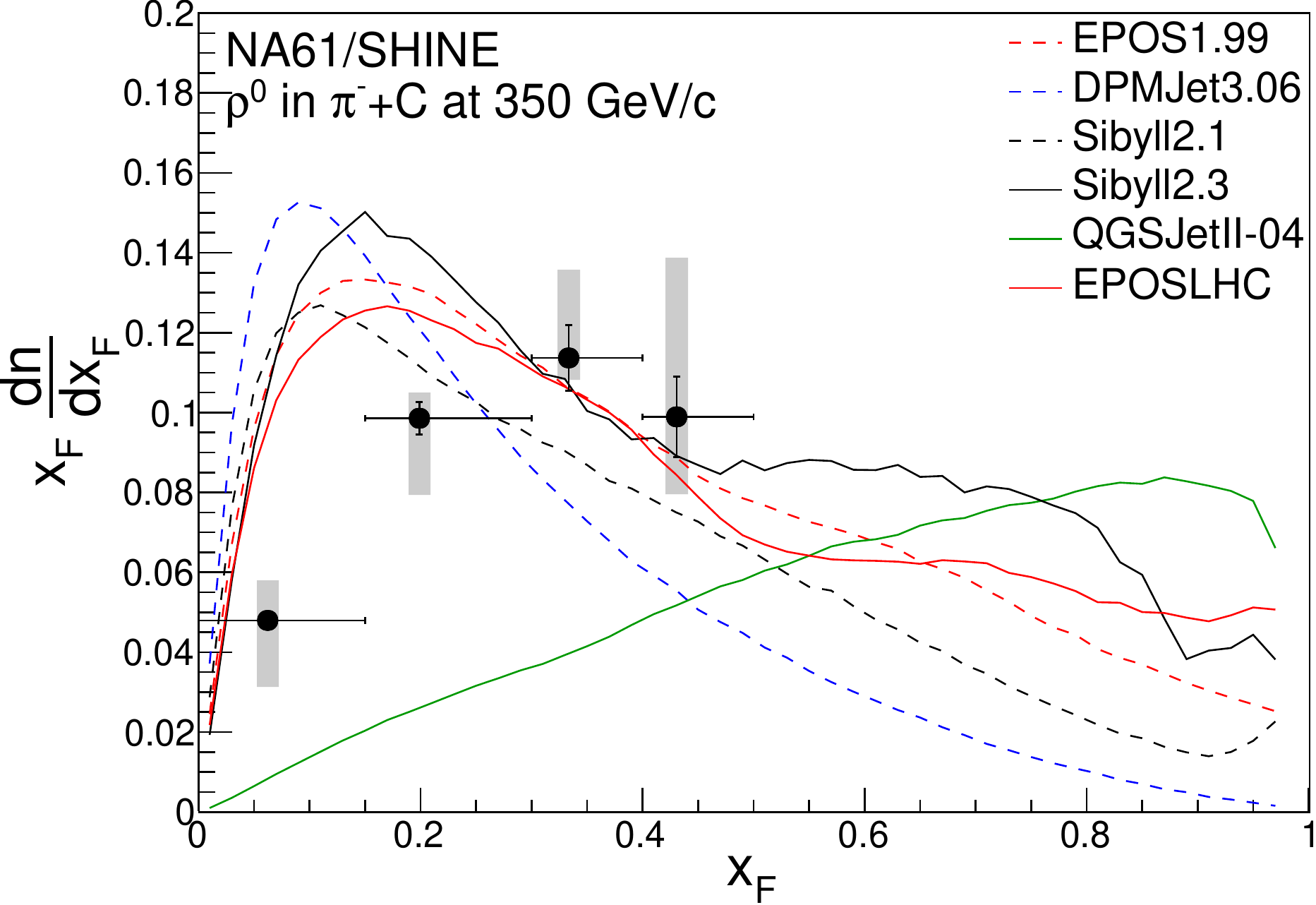}
    \put(15.5,23){}
  \end{overpic}

  \begin{overpic}[clip, rviewport=0 0 1 1,width=0.47\textwidth]{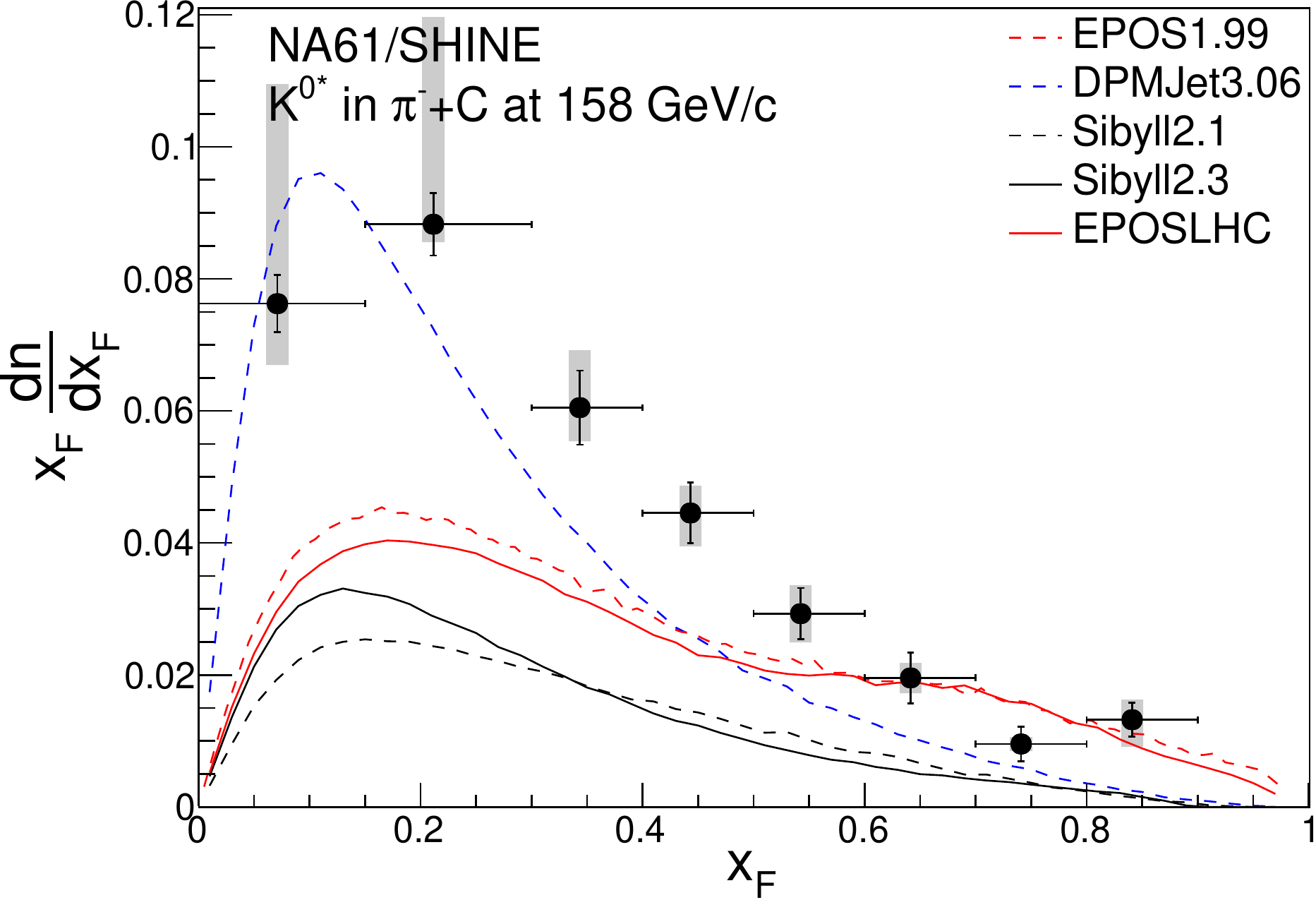}
    \put(60,30){}
  \end{overpic}
  \begin{overpic}[clip, rviewport=0 0 1 1,width=0.47\textwidth]{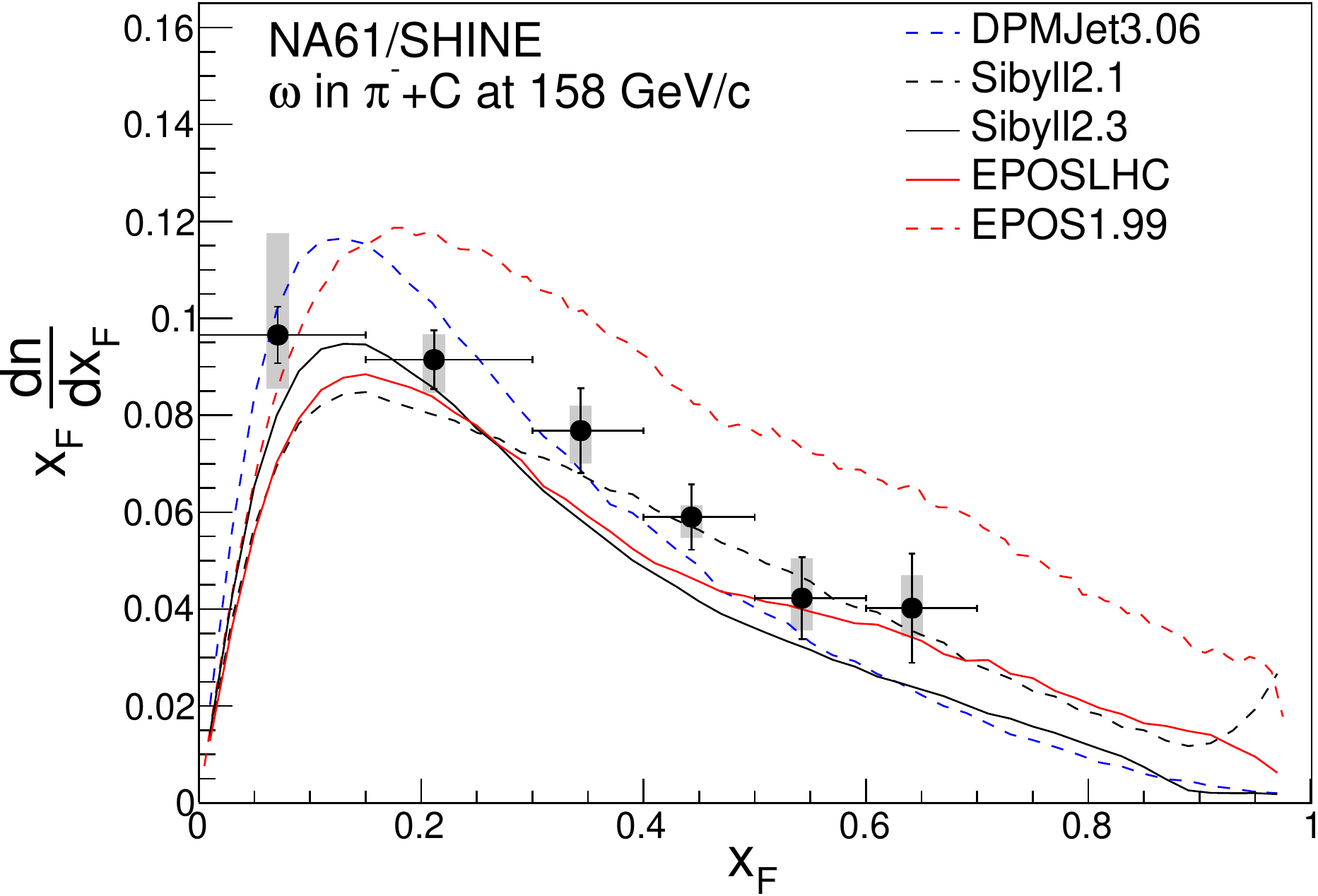}
    \put(15.5,23){}
  \end{overpic}

  \caption{Spectra of \r0, \kstar and $\omega$ as a function of \xF.
    The statistical uncertainties are shown as black bars and the systematic ones as gray bands.}
  \label{fig:resonance:spec}
\end{figure*}

Finally, we have shown the final results of 
the meson resonance analysis which have
already been published in Ref.~\cite{\RhoPaper}.
From the \r0 production spectra shown in~\cref{fig:resonance:spec},
one can see that none of the hadronic models
can describe well the measurements. The small excess of 
\r0 observed with relation to the predictions from simulations
can be relevant to explain the muon deficit in simulations.

%%%%SPEC HADRON

%%%%SPEC VZERO

%%%%SPEC RESONANCE

%%%%%%%%%%%%%%%%%%%%%%%%%%%%%%%%
\section{Acknowledgments}

We would like to thank the CERN EP, BE and EN Departments for the strong support of \NASixtyOne.

This work was supported by the Hungarian Scientific Research Fund (Grants NKFIH 123842–123959), the
J\'anos Bolyai Research Scholarship of the Hungarian Academy of Sciences, the Polish Ministry of Science
and Higher Education (grants 667\slash N-CERN\slash 2010\slash 0, NN 202 48 4339 and NN 202 23 1837), the Polish
National Center for Science (grants 2011\slash 03\slash N\slash ST2\slash 03691, 2013\slash11\slash N\slash ST2\slash03879, 2014\slash 13\slash N\slash ST2\slash02565,
2014\slash14\slash E\slash ST2\slash00018, 2014\slash15\slash B\slash ST2\slash02537 and 2015\slash18\slash M\slash ST2\slash00125, 2015\slash19\slash N\slash ST2 \slash01689,
2016\slash23\slash B\slash ST2\slash00692), the Russian Science Foundation, grant 16-12-10176, the Russian Academy of
Science and the Russian Foundation for Basic Research (grants 08-02-00018, 09-02-00664 and
12-02-91503-CERN), the Ministry of Science and Education of the Russian Federation, grant No.
3.3380.2017\slash4.6, the National Research Nuclear University MEPhI in the framework of the Russian
Academic Excellence Project (contract No. 02.a03.21.0005, 27.08.2013), the Ministry of Education,
Culture, Sports, Science and Technology, Japan, Grant-in-Aid for Scientific Research (grants 18071005,
19034011, 19740162, 20740160 and 20039012), the German Research Foundation (grant GA 1480\slash2-2),
the Bulgarian Nuclear Regulatory Agency and the Joint Institute for Nuclear Research, Dubna (bilateral
contract No. 4418-1-15\slash17), Bulgarian National Science Fund (grant DN08\slash11), Ministry of Education and
Science of the Republic of Serbia (grant OI171002), Swiss Nationalfonds Foundation (grant
200020117913\slash1), ETH Research Grant TH-01 07-3 and the U.S. Department of Energy.

\bibliography{lib}

\end{document}

%% file: TexAbbreviations.tex
\newcommand{\eV}{\ensuremath{\mbox{e\kern-0.1em V}}\xspace}
\newcommand{\EeV}{\ensuremath{\mbox{Ee\kern-0.1em V}}\xspace}
\newcommand{\TeV}{\ensuremath{\mbox{Te\kern-0.1em V}}\xspace}
\newcommand{\PeV}{\ensuremath{\mbox{Pe\kern-0.1em V}}\xspace}
\newcommand{\GeV}{\ensuremath{\mbox{Ge\kern-0.1em V}}\xspace}
\newcommand{\MeV}{\ensuremath{\mbox{Me\kern-0.1em V}}\xspace}
\newcommand{\GeVc}{\ensuremath{\mbox{Ge\kern-0.1em V}\kern-0.1em/\kern-0.05em c}\xspace}
\newcommand{\GeVcc}{\ensuremath{\mbox{Ge\kern-0.1em V}\kern-0.1em/\kern-0.05em c^2}\xspace}
\newcommand{\AGeV}{\ensuremath{A\,\mbox{Ge\kern-0.1em V}}\xspace}
\newcommand{\AGeVc}{\ensuremath{A\,\mbox{Ge\kern-0.1em V}\kern-0.1em/\kern-0.05em c}\xspace}
\newcommand{\MeVc}{\ensuremath{\mbox{Me\kern-0.1em V}\kern-0.1em/\kern-0.05em c}\xspace}
\newcommand{\MeVcc}{\ensuremath{\mbox{Me\kern-0.1em V}\kern-0.1em/\kern-0.05em c^2}\xspace}
\newcommand{\T}{\ensuremath{\mbox{T}}\xspace}

\newcommand{\xF}{\ensuremath{x_\text{F}}\xspace}

\newcommand{\minv}{\ensuremath{m_\text{inv}}\xspace}

\newcommand{\pT}{\ensuremath{p_\text{T}}\xspace}
\newcommand{\pp}{\ensuremath{p}\xspace}

\newcommand{\Epos}{{\scshape Epos}\xspace}
\newcommand{\EposLong}{{\scshape Epos\,1.99}\xspace}

\newcommand{\QGSJetLong}{{\scshape QGSJet\,II-04}\xspace}

\newcommand{\SibyllLong}{{\scshape Sibyll\,2.1}\xspace}
\newcommand{\SibyllNewLong}{{\scshape Sibyll\,2.3}\xspace}
\newcommand{\EposLHCLong}{{\scshape Epos\,LHC}\xspace}

\newcommand{\NASixtyOne}{NA61\slash SHINE\xspace}%this seems to work properly to me. aa
\newcommand{\CernVM}{\textsc{Cern\-\kern-0.05emVM}\xspace}

%% file: ThisAbbreviations.tex
\newcommand{\nmu}{\ensuremath{\mbox{$N_\mu$}}\xspace}

\def \piC{$\pi^-$+C\xspace}
\def \pions{$\pi^\pm$\xspace}
\def \kaons{K$^\pm$\xspace}
\def \proton{p\xspace}
\def \antiproton{$\bar{\text{p}}$\xspace}

\def \protons{p($\bar{\text{p}}$)\xspace}

\def \lamb{$\Lambda$\xspace}
\def \antilamb{$\bar{\Lambda}$\xspace}
\def \lambs{$\Lambda(\bar{\Lambda})$\xspace}
\def \kzeros{K$_\text{S}^{0}$\xspace}
\def \pipi{$\pi^+\pi^-$\xspace}
\def \vzero{$V^0$\xspace}

\def \kstar{K$^{*0}$\xspace}
\def \r0{$\rho^{0}$\xspace}
\def \minvpi{$m_\text{inv}(\pi^+\pi^-)$\xspace}
\def \minv{$m_\text{inv}$\xspace}

\newcommand{\dedx}{\ensuremath{\mbox{\text{d}$E$/\text{d}$x$}}\xspace}

%% file: bib_aliases.tex
\def\NASixtyOnePaper{Abgrall:2014xwa} 

\def\T2KPaper{Abe:2011ks}

\def\QGSJetPaper{Ostapchenko:2010vb}

\def\EposPaper{Pierog:2006qv}
\def\EposLHCPaper{Pierog:2013ria}
\def\SibyllPaper{Ahn:2009wx}

\def\NewSibyllCPaper{Engel:2017icrc}

\def\AugerHASMuonPaper{Aab:2014pza}
\def\AugerMPDPaper{Aab:2014dua}
\def\AugerTopDownPaper{Aab:2016hkv}

\def\AugerXmaxICRC2017Paper{Bellido:2017cgf}

\def\AugerDeltaPaper{Aab:2017cgk}

\def\HiResMIAMuonPaper{AbuZayyad:1999xa}
\def\IceTopMuonPaper{Dembinski:2017zkb}

\def\SugarMuonPaper{Bellido:2018toz}

\def\RhoPaper{Aduszkiewicz:2017anm}

\def\APP16{Prado:2016akv}
\def\APP17{Prado:2017ijs}
\def\IoanaICRC{IoanaICRC2009}

\def\AlexICRC{Herve:2015lra}
\def\RaulICRC{Prado:2017hub}

%% file: RaulRPrado_NA61.bbl
\begin{thebibliography}{27}

\bibitem{Kampert:2012mx}
K.H. Kampert, M.~Unger, Astroparticle Physics \textbf{35}, 660 (2012),
  \texttt{1201.0018}

\bibitem{Engel:2011zzb}
R.~Engel, D.~Heck, T.~Pierog, Annual Review of Nuclear and Particle Science
  \textbf{61}, 467 (2011)

\bibitem{AbuZayyad:1999xa}
T.~Abu-Zayyad et~al. (HiRes/MIA Collaboration), Physical Review Letters
  \textbf{84}, 4276 (2000), \texttt{astro-ph/9911144}

\bibitem{Aab:2014pza}
A.~Aab et~al. (Pierre Auger Collaboration), Physical Review \textbf{D91},
  032003 (2015), \texttt{1408.1421}

\bibitem{Aab:2016hkv}
A.~Aab et~al. (Pierre Auger Collaboration), Physical Review Letters
  \textbf{117}, 192001 (2016), \texttt{1610.08509}

\bibitem{Aab:2014dua}
A.~Aab et~al. (Pierre Auger Collaboration), Physical Review \textbf{D90},
  012012 (2014), \texttt{1407.5919}

\bibitem{Aab:2017cgk}
A.~Aab et~al. (Pierre Auger Collaboration), Physical Review \textbf{D96},
  122003 (2017), \texttt{1710.07249}

\bibitem{Abbasi:2018fkz}
R.U. Abbasi et~al. (Telescope Array Collaboration), Phys. Rev. \textbf{D98},
  022002 (2018), \texttt{1804.03877}

\bibitem{Apel:2017thr}
W.D. Apel et~al. (KASCADE-Grande Collaboration), Astroparticle Physics
  \textbf{95}, 25 (2017)

\bibitem{Dembinski:2017zkb}
H.~Dembinski (IceCube Collaboration), EPJ Web of Conferences \textbf{145},
  01003 (2017)

\bibitem{Bellido:2018toz}
J.A. Bellido, R.W. Clay, N.N. Kalmykov, I.S. Karpikov, G.I. Rubtsov, S.V.
  Troitsky, J.~Ulrichs, Phys. Rev. \textbf{D98}, 023014 (2018),
  \texttt{1803.08662}

\bibitem{Meurer:2005dt}
C.~Meurer, J.~Bluemer, R.~Engel, A.~Haungs, M.~Roth, Czechoslovak Journal of
  Physics \textbf{56}, A211 (2006), \texttt{astro-ph/0512536}

\bibitem{IoanaICRC2009}
{I. C. Maris (for the NA61/SHINE Collaboration)}, Proc. of 31st ICRC  (2009)

\bibitem{Pierog:2006qv}
T.~Pierog, K.~Werner, Physical Review Letters \textbf{101}, 171101 (2008),
  \texttt{astro-ph/0611311}

\bibitem{Drescher:2007hc}
H.J. Drescher, Physical Review \textbf{D77}, 056003 (2008), \texttt{0712.1517}

\bibitem{Abgrall:2014xwa}
N.~Abgrall et~al. (NA61/SHINE Collaboration), Journal of Instrumentation
  \textbf{9}, P06005 (2014), \texttt{arXiv:1401.4699}

\bibitem{Abgrall:2015hmv}
N.~Abgrall et~al. (NA61/SHINE Collaboration), European Physics Journal
  \textbf{C76}, 84 (2016), \texttt{1510.02703}

\bibitem{Aduszkiewicz:2017sei}
A.~Aduszkiewicz et~al. (NA61/SHINE Collaboration), European Physics Journal
  \textbf{C77}, 671 (2017), \texttt{1705.02467}

\bibitem{Abe:2011ks}
K.~Abe et~al. (T2K Collaboration), Nuclear Instruments and Methods in Physics.
  \textbf{A659}, 106 (2011), \texttt{1106.1238}

\bibitem{Prado:2017hub}
{R. R. Prado (for the NA61/SHINE Collaboration)}, Proc. of 35th ICRC  (2017),
  \texttt{1707.07902}

\bibitem{Ahn:2009wx}
E.J. Ahn, R.~Engel, T.K. Gaisser, P.~Lipari, T.~Stanev, Physical Review
  \textbf{D80}, 094003 (2009), \texttt{0906.4113}

\bibitem{Engel:2017icrc}
R.~Engel, F.~Riehn, A.~Fedynitch, T.K. Gaisser, T.~Stanev, Proc. of 35th ICRC
  (2017)

\bibitem{Ostapchenko:2010vb}
S.~Ostapchenko, Physical Review \textbf{D83}, 014018 (2011), \texttt{1010.1869}

\bibitem{Pierog:2013ria}
T.~Pierog, I.~Karpenko, J.M. Katzy, E.~Yatsenko, K.~Werner, Physical Review
  \textbf{C92}, 034906 (2015), \texttt{1306.0121}

\bibitem{Herve:2015lra}
{A. Herve (for the NA61/SHINE Collaboration)}, Proc. of 34th ICRC  (2015),
  \texttt{1509.06586}

\bibitem{PradoHEP2018}
{R. R. Prado (for the NA61/SHINE Collaboration)}, in \emph{7th International
  Conference on High Energy Physics in the LHC Era} (Valparaiso, Chile, 2018)

\bibitem{Aduszkiewicz:2017anm}
A.~Aduszkiewicz et~al. (NA61/SHINE Collaboration), European Physics Journal
  \textbf{C77}, 626 (2017), \texttt{1705.08206}

\end{thebibliography}
